\documentclass[aps,prl,twocolumn,preprintnumbers,amsmath,amssymb,superscriptaddress,floatfix,10pt]{revtex4-2}
\usepackage{amsmath,graphicx}
\usepackage[utf8]{inputenc}
\usepackage[T1]{fontenc}
\usepackage{xcolor}
\usepackage{textcomp}
\usepackage{bm}
\usepackage[normalem]{ulem}

\usepackage{physics}
\usepackage{amsmath} 
\usepackage{tikz}
\usepackage{mathdots}
\usepackage{yhmath}
\usepackage{cancel}
\usepackage{color}
\usepackage{array}
\usepackage{multirow}
\usepackage{amssymb}
\usepackage{gensymb}
\usepackage{tabularx}
\usepackage{extarrows} 
\usepackage{booktabs}
\usetikzlibrary{fadings}
\usetikzlibrary{patterns}
\usetikzlibrary{shadows.blur}
\usetikzlibrary{shapes}

\usepackage{color}
\definecolor{LinkColor}{rgb}{0.75,0.0,0.2}

\usepackage{hyperref}
\hypersetup{
	pdfauthor={good guys},
	pdftitle={good title},
	colorlinks=true,
	citecolor=LinkColor,
	linkcolor=LinkColor,
	urlcolor=LinkColor,
}

\usepackage{listings}
\definecolor{lightgray}{gray}{1}

\usepackage{theorem}

\usepackage{tikz}

\newcommand{\nc}{\newcommand}
\nc{\braoprket}[3]{\langle#1|#2|#3\rangle}
\nc{\opn}[1]{\operatorname{#1}}
\nc{\avg}[1]{\langle#1\rangle}
\nc{\ketbrasame}[1]{|#1\rangle\!\langle#1|}
\nc{\swap}{\opn{SWAP}}
\nc{\E}{\mathbb{E}}
\nc{\Var}{\opn{Var}}
\nc{\dg}{\dagger}

\usepackage[normalem]{ulem}

\begin{document}

\title{Topological Quantum Criticality in Quasiperiodic Cluster Ising Chain}

\author{Sheng Yang}
\affiliation{Institute for Advanced Study in Physics and School of Physics, Zhejiang University, Hangzhou 310058, China}

\author{Hai-Qing Lin}
\email{hqlin@zju.edu.cn}
\affiliation{Institute for Advanced Study in Physics and School of Physics, Zhejiang University, Hangzhou 310058, China}

\author{Xue-Jia Yu}
\email{xuejiayu@eitech.edu.cn}
\affiliation{Eastern Institute of Technology, Ningbo 315200, China}

\begin{abstract}
Topological classifications of quantum critical systems have recently attracted growing interest, as they go beyond the traditional paradigms of condensed matter and statistical physics.
However, such classifications remain largely unexplored in aperiodic environments, particularly under quasiperiodic modulations.
In this Letter, we uncover a family of topological quasiperiodic fixed points with critical scalings intermediate between the clean and infinite-randomness limits.
By exactly solving the quasiperiodic cluster-Ising chain, we demonstrate that all phase boundaries separating quasiperiodically modulated phases are governed by topological Ising-like fixed points unique to strongly modulated quasiperiodic systems.
Although their local bulk critical properties match those of conventional quasiperiodic Ising critical points reported before, these fixed points host robust topological edge degeneracies and exhibit a distinct symmetry-enriched realization of the quasiperiodic Ising criticality, as further supported by complementary lattice simulations.
\end{abstract}

\maketitle

\emph{Introduction.}---Discovering universality classes of phase transitions that go beyond the conventional Landau–Ginzburg–Wilson symmetry-breaking paradigm constitutes a central theme in modern statistical and condensed-matter physics.
For decades, the classification of universality classes was long believed to be uniquely determined by a set of critical exponents, forming the standard framework for investigating phase transitions both theoretically and experimentally~\cite{landau2013statistical,Sachdev_1999,Sondhi1997RMP}.
This belief has been challenged by recent advances in the discovery of topological physics in quantum critical systems~\cite{YU20261,Scaffidi2017PRX,Verresen2018PRL,Verresen2021PRX,Yu2022PRL}.
In particular, symmetry-protected topological edge modes—previously thought to exist only in gapped systems—have been shown to coexist with a gapless bulk at criticality, resulting in topologically distinct realizations of the same universality class even when they share the same local bulk critical exponents, now referred to as topologically nontrivial or symmetry-enriched quantum critical points (QCPs)~\cite{Verresen2021PRX,Duque2021PRB_L,Yu2022PRL,verresen2020topologyedgestatessurvive,Ye2022SciPost,Mondal2023PRB,Yu2024PRB,Choi2024PRB,Zhong2024PRA,Saran2024PRB_L,Li2025PRB_L,Zhou2025CP,Cardoso2025PRB,Zhong2025PRB,tan2025exploringnontrivialtopologyquantum,chou2025ptsymmetryenrichednonunitarycriticality,Rey2025PRB,Moy2025SciPost,guo2025generalizedlihaldanecorrespondencecritical,prembabu2025noninvertibleinterfacessymmetryenrichedcritical,deng2025anomalousdynamicalscalingtopological,Jia2025PRL}, and more generally to gapless symmetry-protected topological (SPT) states~\cite{Scaffidi2017PRX,Verresen2021PRX,Thorngren2021PRB,Hidaka2022PRB,Wen2023PRB,Yu2024PRL,Su2024PRB,Zhang2024PRA,Li2024SciPost,wen2024topologicalholographyfermions,huang2024fermionicquantumcriticalitylens,Li2025SciPost,Yang2025CP,yang2025deconfinedcriticalityintrinsicallygapless,wen2025stringcondensationtopologicalholography,yu2025gaplesssymmetryprotectedtopologicalstates,Wen2025PRB,Flores2025PRL,Huang2025SciPost,wen2025topologicalholography21dgapped,prembabu2025multicriticalitypurelygaplessspt,banerjee2025entanglementspectrumgaplesstopological}. The corresponding edge state can also be systematically constructed within the framework of conformal field theory (CFT)~\cite{KevinGraham_2004,frohlich2004PRL,Furuya2017PRL,LECHEMINANT2015510,Fukusumi2021PRB,Wu2026PRL,Okada2024PRL,Runkel2020JHEP,smith2021boundarystatesanomaloussymmetries}.
This conceptual extension not only generalizes the notion of topology to gapless quantum critical systems, but also reveals fundamentally new phenomena beyond gapped topological phases, including nontrivial conformal boundary conditions~\cite{FREDENHAGEN2003436,WATTS2012177,Affleck1991PRL,Friedan2004PRL,Yu2022PRL,Parker2018PRB}, algebraically localized edge modes~\cite{Verresen2021PRX,Yang2025CP}, universal bulk-boundary correspondence~\cite{Yu2024PRL,Zhong2025PRB,guo2025generalizedlihaldanecorrespondencecritical,banerjee2025entanglementspectrumgaplesstopological}, anomalous dynamical scaling behaviors~\cite{Parker2019PRL,deng2025anomalousdynamicalscalingtopological}, and intrinsically gapless topological states~\cite{Thorngren2021PRB,Li2025SciPost,Zhang2024PRA,yang2025deconfinedcriticalityintrinsicallygapless}. 

A natural and pressing question is how such topological quantum criticality manifests in realistic settings, where imperfections and inhomogeneities are unavoidable and may profoundly modify the universality classes of phase transitions~\cite{Vojta2019AnnualReview,Fisher1992PRL,Fisher1995PRB,Motrunich2000PRB,Luck1993JSP}.
A particularly intriguing form of imperfection in many modern experimental platforms—including superconducting qubit arrays~\cite{Shi2023PRL,huang2025exactquantumcriticalstates}, moiré materials~\cite{Uri2023Nature,Lai2025NM}, and cold atoms in bichromatic optical lattices~\cite{Bordia2017PRX,Deissler2010NP,Luschen2017PRX}—is intrinsic inhomogeneity arising from quasiperiodic (QP) modulations of the couplings.
These modulations can be regarded as a form of correlated randomness and are therefore expected to realize universality classes absent in both clean and randomly disordered systems. Nevertheless, over the past decades, studies of phase transitions in QP systems have focused primarily on localization-delocalization phenomena~\cite{aubry1980analyticity,Harper_1955,Das1988PRL,Ringot2000PRL,Evers2008RMP,Biddle2009PRA,Biddle2010PRL,Iyer2013PRB,Ganeshan2015PRL,Khemani2017PRL,Yao2019PRL,Longhi2019PRL,Doggen2019PRB,Nicolas2019SciPost,Wang2020PRL,Lv2022PRA,Gonifmmode2023PRL,Yang2024PRA}, particularly the pioneering works of Kohmoto and collaborators, which revealed new localization transitions, scaling behavior, and critical (multifractal) spectra~\cite{Kohmoto1983PRL,Kohmoto1987PRB,Kohmoto1983PRL_b,Hiramoto1989PRL}.
By contrast, \emph{ground-state} quantum phase transitions in such systems—despite recent progress~\cite{OliveiraFilho_2012,Deguchi2012NM,Chandran2017PRX,Crowley2018PRL,Divakaran2018PRE,Crowley2019PRB,Agrawal2020NC,Agrawal2020PRL,Cookmeyer2020PRB,Revathy_2020,Khosravian2021PRR,Crowley_2022,yeo2022nonpowerlawuniversalscalingincommensurate,Shkolnik2023PRB,Karcher2024PRB,Gonifmmode2024PRB,Gonifmmode2024PRB_b,Vongkovit2024PRB_L,zhang2025magneticordernovelquantum}—are far less explored.
Previous works have argued that clean critical points are perturbatively more stable against QP than random perturbations, based on the general stability criterion proposed by Luck~\cite{Luck1993JSP}.
However, a more fundamental and intriguing open question concerns how strong QP modulations interplay with topology and whether they can give rise to QCPs that are genuinely distinct from those in conventional homogeneous systems.

In this Letter, we fill this gap by uncovering a family of topological quasiperiodic fixed points unique to QP modulated systems.
By analytically establishing the ground-state phase diagram of the QP cluster-Ising spin chain, we first show that strong QP modulations induce an unconventional topological phase with area-law entanglement despite a vanishing bulk energy gap.
More importantly, we unambiguously demonstrate that, although all phase boundaries separating quasiperiodically modulated phases lack CFT descriptions, they can still be enriched by certain global symmetries, giving rise to topological critical points with robust topological edge degeneracies and critical scalings intermediate between the clean and infinite-randomness limits.
Consequently, the QP cluster-Ising transition identified here is topologically distinct from the previously recognized QP-Ising transition in Ref.~\cite{Crowley2018PRL}, as they cannot be smoothly connected without encountering a multicritical point or explicitly breaking the protecting symmetries.

\begin{figure}[t]
  \centering
  \includegraphics[width=\linewidth]{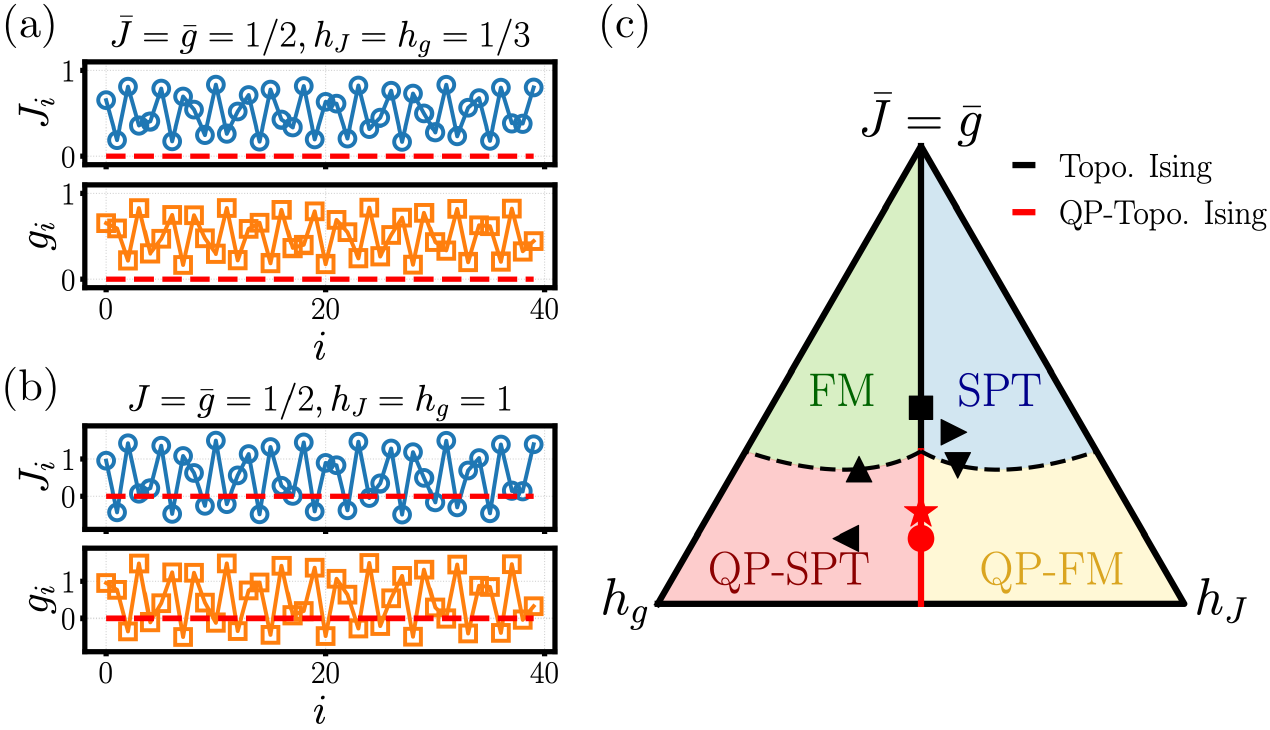}
  \caption{(a-b) Spatial profiles of couplings $J_{i}$ and $g_{i}$ for two representative points marked in (c) (black square and red star). Parameters: fixed $\phi_{1} = 1/10$ and $\phi_{2} = 1$. (c) Ground-state phase diagram of the QP cluster-Ising model containing FM, QP-FM, SPT, and QP-SPT phases. The vertical black and red lines represent the topological Ising and QP-Topological Ising critical lines, respectively. Symbols indicate points analyzed in later figures.}
  \label{fig:phase_diagram}
\end{figure}

\emph{Model.}---We consider a spin-$1/2$ cluster-Ising chain~\cite{Verresen2018PRL,Verresen2021PRX,Yu2022PRL} with QP modulations,
\begin{equation}
  \label{eq:model}
  H_\text{CI}^\text{QP} = - \sum_{i} J_{i} \sigma_{i}^{x} \sigma_{i+1}^{x} - \sum_{i} g_{i} \sigma_{i}^{x} \sigma_{i+1}^{z} \sigma_{i+2}^{x} \, ,
\end{equation}
which preserves a $\mathbb{Z}_2$ spin-flip symmetry generated by $P=\prod_i \sigma_i^z$ and a time-reversal symmetry $\mathbb{Z}_2^{T}$ generated by $T = K$ the complex conjugation.
Together they form the protecting symmetry $G = \mathbb{Z}_{2} \times \mathbb{Z}_{2}^{T}$.
In particular, the QP modulated couplings are chosen as [our conclusions apply to generic continuous strong QP modulations for which the couplings vanish at certain lattice sites~\cite{Crowley_2022}, as illustrated in Fig.~\ref{fig:phase_diagram}(b)]
\begin{align}
  \label{eq:couplings}
  J_{i} & = \bar{J} + h_{J} \cos{[Q(i+1/2)+\phi_{1}]} \, , \\ 
  g_{i} & = \bar{g} + h_{g} \cos{[Qi+\phi_{1}+\phi_{2}]} \, .
\end{align}
Here $\sigma_i^{x,y,z}$ denote Pauli matrices at site $i$, and $J_i$ ($g_i$) are the QP modulated Ising (cluster) couplings with modulation strengths $h_J$ ($h_g$).
The modulation is quasiperiodic when the wavelength $2\pi/Q$ is an irrational multiple of the lattice constant $a=1$. 
In this work, we set $Q/(2\pi) = \tau_\text{G}$ with $\tau_\text{G} = (\sqrt{5} + 1) /2$ the Golden ratio.
In numerical simulations with periodic boundary conditions (PBCs), we employ rational approximants $Q/(2\pi)=p_i/q_i=F_{i+2}/F_{i+1}$, where $F_i$ are Fibonacci numbers. This approximation can be viewed as a periodic spin chain with $q_i$ local degrees of freedom per unit cell. At criticality, the low-energy theory is expected to be described by a multi-component or orbifold CFT, $(\mathrm{Ising\ CFT})^{q_i}/\mathbb{Z}_{q_i}$~\cite{francesco2012conformal} for sufficiently large system sizes commensurate with $q_i$, and by a defect CFT when the system size is incommensurate~\cite{ALCARAZ1988280,PETKOVA2001157,Watanabe2023JMP,Hu2023PRB,Roy2022PRL,Henning2022SciPost,northe2026entanglementtopologicaldefectsreconciling}. However, since our work focuses on QP quantum criticality, the scaling analysis is restricted to length scales $r < q_i$, where such emergent CFT or defect-CFT descriptions do not arise.
Accordingly, the system size is set to $N=2q_{i}$, which ensures that the modulation is commensurate with the lattice periodicity.
The associated period $q_i$ thus provides a sequence of finite length scales for scaling analysis. 
Hereafter, we denote $q_{i}$ simply as $q$ for brevity.
For open boundary conditions (OBCs), we fix $Q/(2\pi)$ directly to $\tau_\text{G}$.
Our results are not restricted to this choice and apply to a broad class of irrational wave vectors [See Sec.~III of the Supplemental Material (SM) for details].
The phases $\phi_{1}$ and $\phi_{2}$ shift the modulations relative to the lattice;
Most of the calculations presented below involve averaging over different samples of $\phi_{1/2}$ to obtain reliable critical scaling while reducing the system-size requirements; for wandering, however, fixed phases are used as all configurations are effectively represented within a single large system~\cite{Fisher1992PRL,Miranda_2005,Vojta_2006}.

\emph{Ground-state phase diagram.}---In this work, we focus on the \emph{ground-state} properties of QP systems, rather than on localization or dynamical phenomena that are more commonly studied in the literature.
The QP cluster-Ising model defined in Eq.~\eqref{eq:model} is exactly solvable via a Jordan–Wigner transformation, enabling us to analytically establish its ground-state phase diagram~\cite{Crowley2018PRL,Crowley_2022} (also see SM Sec.~I), shown in Fig.~\ref{fig:phase_diagram}(c).

In the weak-modulation regime ($\bar{J}=\bar{g}>h_J,h_g$), the QP perturbation is irrelevant, and the symmetry-enriched phase boundaries (see the End Matter for a brief review) between the ferromagnetic (FM) and the $\mathbb{Z}_2 \times \mathbb{Z}_2^{T}$ cluster SPT phase are preserved.
In contrast, the strong-modulation regime ($\bar{J}=\bar{g}<h_J,h_g$) hosts two distinct gapless phases separated by three phase boundaries.
One is a QP-FM phase that has been studied previously~\cite{Chandran2017PRX,Crowley_2022}, while the other is an unconventional SPT phase that exhibits area-law entanglement scaling despite a vanishing bulk energy gap—a feature unique to QP systems—which we refer to as the QP-SPT phase (see End Matter for details).
The central question addressed in this work is whether symmetry-enriched quantum criticality persists under strong QP modulation, and whether such topological quantum criticality is intrinsic to QP settings.

\emph{Quasiperiodic Ising universality class.}---We now turn to the phase transitions in the presence of strong QP modulation.
In the main text, we focus on a representative strongly modulated critical point at $\bar{J}=\bar{g}=1/2$ and $h_J=h_g=1$ [the red star in Fig.~\ref{fig:phase_diagram}(c)], which separates the QP-SPT and QP-FM phases.
For comparison, we also analyze the weakly modulated Ising transition at $\bar{J}=\bar{g}=1/2$ and $h_J=h_g=1/3$ [the black square in Fig.~\ref{fig:phase_diagram}(c)], where the QP modulation is irrelevant~\cite{Luck1993JSP,Crowley2018PRL} and belongs to clean symmetry-enriched universality.
Throughout this work, we focus on mean-averaged physical observables (see End Matter), which exhibit scaling behaviors analogous to those of homogeneous systems but with modified exponent values.
We first demonstrate that the QP critical points identified here are intermediate between those of clean and infinite-randomness fixed points.

\begin{figure}[t]
  \centering
  \includegraphics[width=\linewidth]{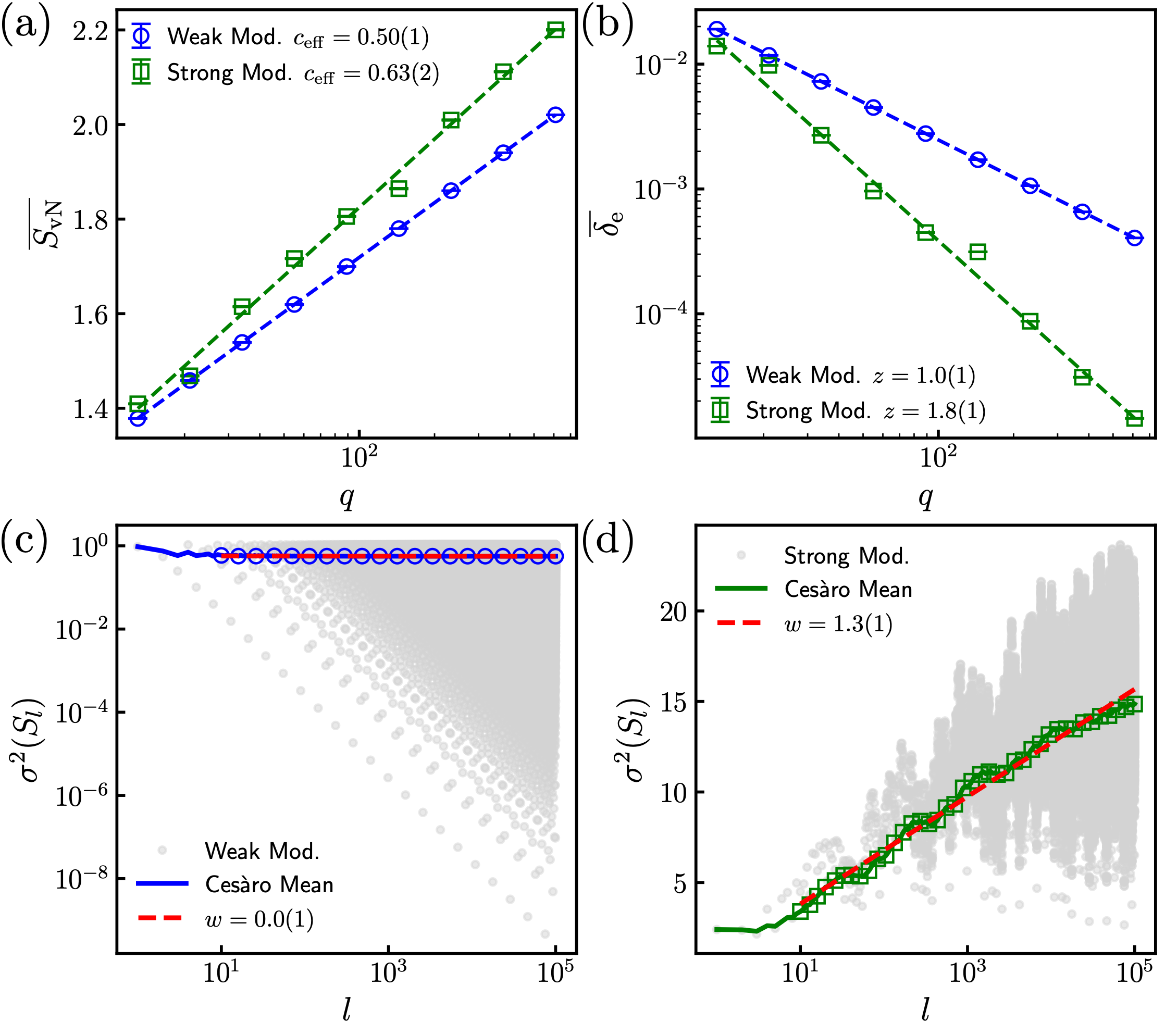}
  \caption{(a) Logarithmic-law behavior of the half-chain  entanglement entropy $\overline{S_{\rm vN}} = \frac{c_\text{eff}}{3} \ln q + {\rm const}$. (b) Energy gap $\overline{\delta_\text{e}}$ versus $q$ for strongly and weakly modulated QCP. The dashed lines represent fits to the scaling form $\overline{\delta_\text{e}}(q) \sim 1/q^{z}$. (c-d) Variance of the wandering $\sigma^{2}(S_{l})$ as a function of $l$ for weakly and strongly modulated QCPs, respectively. The blue and green lines are Ces\`aro means and the red dashed lines are fits of $\sigma^{2}(S_{l}) \propto w \log(l)$; the markers represent the data used in fits. Parameters: $\bar{J} = \bar{g} = 1/2, h_{J} = h_{g} = 1$ [the red star in Fig.~\ref{fig:phase_diagram}(c)] for strongly modulated QCP and $\bar{J} = \bar{g} = 1/2, h_{J} = h_{g} = 1/3$ [the black square in Fig.~\ref{fig:phase_diagram}(c)] for the weakly modulated QCP. The parameter $q$ ranges from $13$ to $610$ under PBC for (a) and (b). In (c) and (d), we have fixed $\phi_{1} = 1/10$ and $\phi_{2} = 1$.}
  \label{fig:qcp_compare}
\end{figure}

To examine whether QP modulation is sufficient to drive the critical point of the cluster-Ising model into the infinite-randomness regime, we investigate the entanglement properties at both critical points through finite-size scaling of the averaged half-chain entanglement entropy $\overline{S_{\mathrm{vN}}}$, as shown in Fig.~\ref{fig:qcp_compare}(a).
At both transitions, $\overline{S_{\mathrm{vN}}}$ grows logarithmically with system size but with distinct prefactors.
This indicates that, despite the absence of conformal invariance at the strongly modulated critical point, the averaged entanglement entropy nevertheless follows the standard (1+1)-dimensional CFT scaling form known from clean systems~\cite{Calabrese_2004}, albeit with an effective central charge $c_{\mathrm{eff}}=0.63(2)$, which differs from both the clean ($c=1/2$) and the infinite randomness Ising QCP ($c_{\mathrm{eff}}=\frac{1}{2}\ln 2$~\cite{Duque2021PRB_L}).
The salient features of this QP fixed point are further corroborated by the finite-size scaling of the bulk energy gap [Fig.~\ref{fig:qcp_compare}(b)], which exhibits algebraic scaling $\overline{\delta_\text{e}} \sim q^{-z}$ at both the weakly (blue curve) and strongly modulated (green curve) transition points, with dynamical critical exponents $z \approx 1.0$ and $z=1.8(1)$, respectively. 
Note that the latter is also consistent with the value reported for the strongly modulated QP transverse-field Ising chain~\cite{Crowley2018PRL,Crowley_2022}, which also constitutes the initial evidence that the bulk critical behavior of the QP cluster-Ising critical point we found matches that of the QP trivial Ising one as detailed further below. 
This behavior indicates that, although strong QP modulation in our context breaks conformal invariance at criticality, it is insufficient to drive the transition to an infinite-randomness fixed point (IRFP) that is characterized by exponentially fast gap closing ($z=\infty$); it also rules out the recently proposed infinite-quasiperiodic transition scenario~\cite{Agrawal2020NC,Agrawal2020PRL} (see End Matter for more discussions).
Moreover, infinite-randomness physics is typically accompanied by qualitatively distinct scaling of the typical and arithmetic averaged spin correlation functions—exponential decay for the former and algebraic decay for the latter (see SM Sec.~II and the following section).
By contrast, at the QP-Ising transition, no such distinction is observed: the typical and arithmetic averaged spin correlations exhibit the same qualitative power-law scaling, providing further evidence against infinite-randomness criticality.

Another way to substantiate the intermediate universality unique to QP systems is through an analysis of the wandering~\cite{Luck1993JSP}, which captures the leading effect of QP modulation on the critical cluster-Ising chain.
Specifically, the wandering $S_l(j)=\sum_{i=j}^{j+l-1} \ln\left(\left|J_i/g_i\right|\right)$ quantifies fluctuations of the reduced coupling $J_i/g_i$ summed over a segment of length $l$.
More importantly, the scaling behavior of the wandering variance $\sigma^{2}(S_l)=[S_l(j)^2]-[S_l(j)]^{2}$, where $[\,\cdot\,]$ means the average over the site position $j$, serves as a diagnostic of different types of disordered criticality~\cite{Torquato2016PRE,Crowley_2022}.
In $(1+1)$ dimensions, $\sigma^{2}(S_l)$ saturates to a constant at a clean critical point, while it grows linearly with $l$ at an uncorrelated disordered critical point. By contrast, at a QP critical point, $\sigma^{2}(S_l)$ exhibits logarithmic scaling, $\sigma^{2}(S_l)\sim w\ln l$, with a nonvanishing coefficient $w$, known as the logarithmic wandering coefficient~\cite{Crowley2018PRL}.
In practice, we characterize the scaling behavior of $\sigma^{2}(S_l)$ at both critical points using the Ces\`aro mean, $[\sigma^{2}(S_l)]_{\mathrm{Ces\grave{a}ro}}=\frac{1}{l}\sum_{l'=1}^{l}\sigma^{2}(S_{l'})$.
As shown in Fig.~\ref{fig:qcp_compare}(c) and Fig.~\ref{fig:qcp_compare}(d), $[\sigma^{2}(S_l)]_{\mathrm{Ces\grave{a}ro}}$ saturates to a constant at the weakly modulated critical point [blue curve in Fig.~\ref{fig:qcp_compare}(c)], whereas it exhibits logarithmic growth only at the strongly modulated critical point, with a coefficient $w\simeq1.3(1)$ [green curve in Fig.~\ref{fig:qcp_compare}(d)], consistent with analytical results for the QP transverse-field Ising chain~\cite{Crowley_2022}.
We therefore conclusively demonstrate that the strongly modulated critical point is controlled by a QP-Ising fixed point intermediate between the clean and infinite-randomness limits.

\emph{Topological features at quasiperiodic transition}---Recent advances~\cite{YU20261,Verresen2018PRL,Yu2022PRL} have demonstrated that two QCPs can be further distinguished by their topological properties even when they share the same local bulk critical exponents and are therefore controlled by distinct fixed points, as they cannot be smoothly connected without either crossing an additional phase transition or breaking the protecting symmetries.
It is therefore necessary to determine whether the QP Ising fixed point found here is topologically distinct from the critical point of the QP transverse-field Ising chain reported in the literature~\cite{Crowley2018PRL,Crowley_2022}.

\begin{figure}[t]
  \centering
  \includegraphics[width=\linewidth]{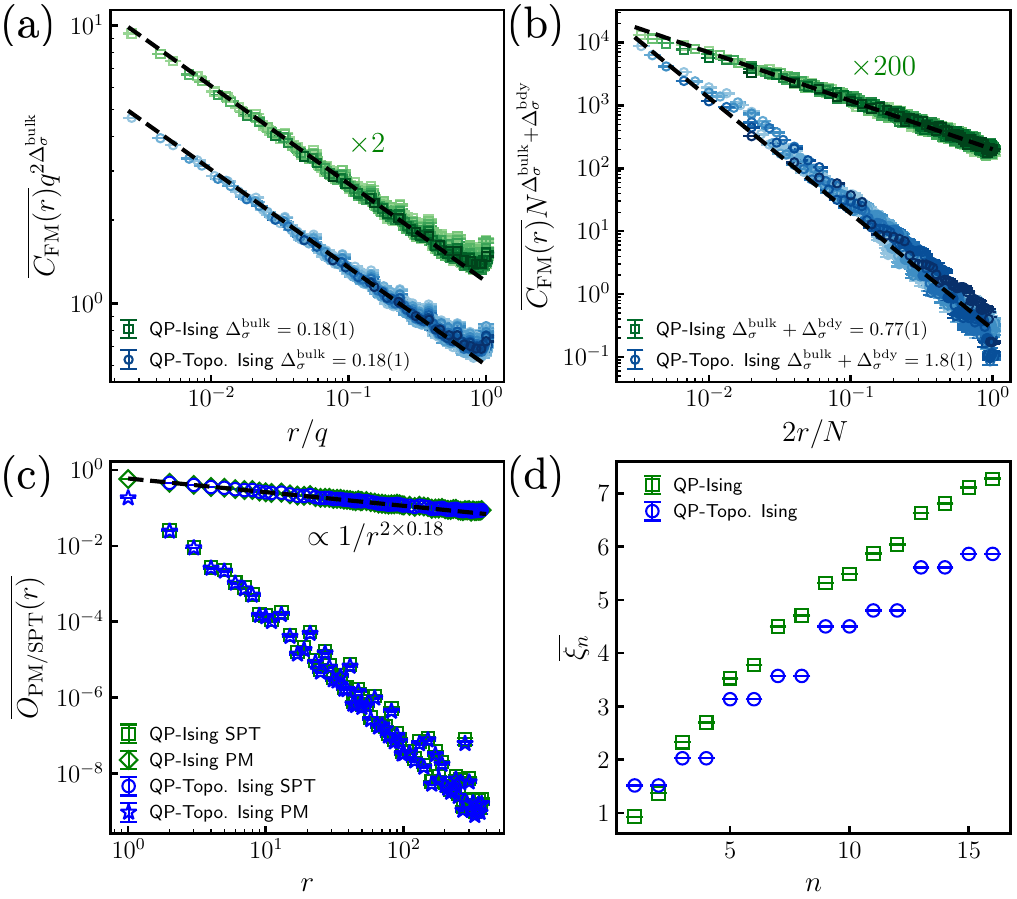}
  \caption{Data collapse of (a) bulk-bulk and (b) boundary-bulk connected correlation functions $\overline{C_\text{FM}(r)}$ for QP-Topological Ising and QP-Ising transitions. The dashed lines are $\propto 1/(r/q)^{2\Delta_{\sigma}^\text{bulk}}$ or $\propto 1/(2r/N)^{\Delta_{\sigma}^\text{bulk} + \Delta_{\sigma}^\text{bdy}}$ for comparison with the data collapse. To achieve a better visualization, we have multiplied an additional global factor for the correlations of QP-Ising as indicated by ``$\times 2$'' and ``$\times 200$''. (c) The power-law behaviors of SPT/PM string order; the corresponding scaling dimensions are extracted via data collapse shown in SM Sec.~II\,. (d) The bulk many-body entanglement spectrum $\overline{\xi_{n}}$. Parameters: $\bar{J} = \bar{g} = 1/2, h_{J} = h_{g} = 1$. The parameter $q$ ranges from $13$ to $377$ under PBC for (a) and $N$ ranges from $100$ to $600$ under OBC for (b). In (c) and (d), we have chosen $q = 377$ and $610$, respectively, under PBC.}
  \label{fig:ising_ci}
\end{figure}

A key hallmark of topologically nontrivial QCPs is that their local bulk properties are indistinguishable from those of the trivial critical point, while nontrivial topological features manifest on nonlocal string operators, entanglement-spectrum degeneracies, and boundary properties.
To demonstrate this in our setting, we compare various properties at the transition points of the QP cluster-Ising chain and the QP transverse-field Ising chain~\cite{Crowley2018PRL,Crowley_2022}. 
Specifically, we first examine the averaged bulk-bulk spin correlation function $\overline{C_{\mathrm{FM}}(r)}$ under PBCs at criticality, as shown in Fig.~\ref{fig:ising_ci}(a).
In both cases, the rescaled correlator $q^{2\Delta_{\sigma}^{\mathrm{bulk}}}\overline{C_{\mathrm{FM}}(r)}$ collapses onto a single curve when plotted as a function of the fractional separation $r/q$, with a common bulk scaling dimension $\Delta_{\sigma}^{\mathrm{bulk}}=0.18(1)$ (see SM Sec.~II for the results of the bulk-bulk energy-energy correlation).
By contrast, the boundary critical behaviors of the two QP Ising fixed points is clearly distinguished by the averaged boundary-bulk spin correlation function [see Fig.~\ref{fig:ising_ci}(b)], which both exhibits power-law decay of the form $r^{-(\Delta_{\sigma}^{\mathrm{bulk}}+\Delta_{\sigma}^{\mathrm{bdy}})}$ but with significantly different boundary scaling dimensions:
$\Delta_{\sigma}^{\mathrm{bdy}}=0.59(2)$ for the QP-Ising and $\Delta_{\sigma}^{\mathrm{bdy}}=1.6(1)$ for the QP cluster-Ising fixed point (see SM Sec.~II for the results of the boundary-bulk energy-energy correlation).
Taken together, while the QP cluster-Ising fixed point identified here shares indistinguishable local bulk critical properties with the QP Ising critical point proposed in Ref.~\cite{Crowley2018PRL,Crowley_2022}, it exhibits fundamentally different boundary behaviors, which can be attributed to nontrivial topological edge degeneracy, as discussed below.
It is noted that the concrete values of the exponents extracted here also support that the QP Ising transition is different from the clean Ising or IRFP (see Table~\ref{tab:scaling_comparison} for a detailed comparison).

\begin{table}[b]
  \caption{\label{tab:scaling_comparison} Comparison of bulk $\Delta_{\sigma}^\text{bulk}$ and boundary $\Delta_{\sigma}^\text{bdy}$ scaling dimensions of the spin operator across three cases: Ising, IRFP, and QP-Ising. $\mathbb{Z}_{2}^{T}$ denotes the time-reversal symmetry and $\tau_\text{G} = (1 + \sqrt{5}) / 2$.}
  \begin{ruledtabular}
    \begin{tabular}{lcccc}
    & \multicolumn{2}{c}{Trivial} & \multicolumn{2}{c}{Enriched by $\mathbb{Z}_{2}^{T}$} \\
    \cline{2-3} \cline{4-5}
    System Type & $\Delta_{\sigma}^{\text{bulk}}$ & $\Delta_{\sigma}^{\text{bdy}}$ & $\Delta_{\sigma}^{\text{bulk}}$ & $\Delta_{\sigma}^{\text{bdy}}$ \\
    \hline
    Clean Ising~\cite{Yu2022PRL} & $1/8$ & $1/2$ & $1/8$ & $2$ \\
    IRFP~\cite{Duque2021PRB_L} & $1-\tau_\text{G}/2$ & $1/2$ & $1-\tau_\text{G}/2$ & $0$ \\
    QP-Ising & $0.18(1)$ & $0.59(2)$ & $0.18(1)$ & $1.6(1)$ \\
    \end{tabular}
  \end{ruledtabular}
\end{table}

To further elucidate the topological distinction between the two QP critical points, we note that although conformal symmetry is absent under strong QP modulation, both QP Ising fixed points—analogously to the Ising CFT—have nonlocal $\mu$ operators that carry distinct time-reversal symmetry charges, $T\mu T=\pm\mu$, providing a definite bulk topological invariant.
Here, $\mu$ denotes the leading $\mathbb{Z}_2$ symmetry-flux operator, identified by the slowest decay of its two-point function.
On the lattice, these operators take the same form of nonlocal string operators as in the clean Ising model, but with modified scaling dimensions—their averaged two-point functions are $\overline{O_{\mathrm{PM}}(r)}$ for the QP Ising transition and $\overline{O_{\mathrm{SPT}}(r)}$ for the QP cluster-Ising transition—Indeed, the nearby QP modulated symmetric phases are still probed by the same order parameters.
This topological invariant cannot change under continuous $\mathbb{Z}_{2} \times \mathbb{Z}_{2}^{T}$-preserving deformations that retain QP-Ising criticality.
Opposite time-reversal charges therefore distinguish different symmetry-enriched realizations of QP-Ising criticality, leading to topologically distinct QP Ising fixed points without CFT descriptions.

These predictions are tested numerically.
As shown in Fig.~\ref{fig:ising_ci}(c), we compute the averaged string operators $\overline{O_{\mathrm{PM}}(r)}$ and $\overline{O_{\mathrm{SPT}}(r)}$ at both transitions.
At the QP cluster-Ising critical point, $\overline{O_{\mathrm{SPT}}(r)}$ exhibits a slower algebraic decay than $\overline{O_{\mathrm{PM}}(r)}$, whereas the situation is reversed at the QP Ising critical point.
These results imply that the two QP Ising critical points are topologically distinct, exhibiting qualitatively different condensation behaviors of the nonlocal $\mu$ operators.
At the QP transverse-field Ising critical point, the condensation of the $\mathbb{Z}_2^T$-neutral $\mu$ operator preserves the $\mathbb{Z}_2$ symmetry near the boundary.
In contrast, at the QP cluster-Ising critical point, condensation of the $\mathbb{Z}_2^T$-charged $\mu$ operator is prohibited, stabilizing a boundary condition with spontaneously broken $\mathbb{Z}_2$ symmetry and an associated twofold topological degeneracy.
As a result, the two transitions exhibit distinct boundary critical behaviors, in agreement with the numerical results shown in Fig.~\ref{fig:ising_ci}(b).
Complementarily, we examine the averaged bulk entanglement spectrum $\overline{\xi_n}$ at both transition points [Fig.~\ref{fig:ising_ci}(d)], observing a robust twofold topological degeneracy at the QP cluster-Ising QCP (blue dots) (see also SM Sec.~II for its robustness against symmetry-preserving perturbations), whereas no such degeneracy is present at the conventional QP Ising QCP (green squares).
Taken together, these results uncover that time-reversal symmetry can enrich QP critical points even in the absence of conformal symmetry, yielding a symmetry-enriched realization of QP-Ising criticality that is topologically distinct from the conventional one, as summarized in Table~\ref{tab:scaling_comparison}.

\emph{Discussion and concluding remarks.}---In SM Sec.~II and IV, we further examine additional representative points along the QP-SPT to QP-FM phase boundary, which exhibit scaling exponents closely matching those discussed in the main text, suggesting that the entire phase boundary realizes the same topologically nontrivial QP-Ising criticality.
Moreover, we show that quasiperiodic and topological characteristics persist qualitatively—albeit with different scaling exponents—at selected points along the other two phase boundaries that separate the QP-SPT and FM phases, as well as the QP-FM and cluster SPT phases.
Therefore, we believe that phase boundaries between SPT and symmetry-breaking phases, when at least one phase is strongly quasiperiodically modulated, are controlled by a continuous line of topological QP fixed points.
Numerical exploration of such QP fixed points faces substantial challenges, including the strong-disorder renormalization-group methods suitable for IRFPs, as well as tensor-network and quantum Monte Carlo simulations that require demanding computational resources, highlighting the advantage of the exact solution used here.

In summary, we reveal a topological Ising-like criticality in strongly QP-modulated systems.
By exactly solving the QP cluster-Ising chain, we establish the ground-state phase diagram and uncover an unconventional QP-SPT phase in which gapped SPT order coexists with vanishing bulk energy gap—a phenomenon distinctive of strong quasiperiodicity.
More importantly, we unambiguously demonstrate that global symmetry can further enrich strongly QP-modulated phase boundaries that lack a CFT description, giving rise to topological QP critical points with critical scalings intermediate between the clean and infinite-randomness limits, as supported by complementary lattice simulations.
The measured local bulk critical scaling agrees with that of the conventional QP-Ising transition within numerical accuracy, while the leading symmetry-flux operator carries a different time-reversal charge.
The two transitions thus exhibit distinct symmetry-enriched realizations of the QP-Ising universality under the $\mathbb{Z}_2 \times \mathbb{Z}_2^T$ protecting symmetry.
This distinction is preserved under symmetry-preserving deformations that retain the QP-Ising criticality.

In the near future, it would be particularly interesting to explore the dynamical and localization properties of symmetry-enriched QP fixed points, as topology and QP modulation are each known to have distinct and profound effects on these phenomena.
It is also worthwhile to develop efficient algorithms for studying QP criticality in the presence of interactions.
From an experimental perspective, the cluster-Ising chain has recently been realized in ultracold-atom~\cite{Petiziol2021PRL,Sylvain2019Science} and superconducting qubit quantum platforms~\cite{tan2025exploringnontrivialtopologyquantum}, where QP couplings can also be conveniently engineered and controlled within the same settings.

\textit{Acknowledgement}:
We thank Romain Vasseur, Yuxuan Guo, Wen-Yuan Liu, Xiang-Ping Jiang, and Kaiyuan Cao for helpful discussions.
X.-J. Yu was supported by the National Natural Science Foundation of China (Grant No.12405034) and a start-up grant from Eastern Institute of Technology, Ningbo. 
This work is also supported by MOST 2022YFA1402701.
The work of S.Y. is supported by China Postdoctoral Science Foundation (Certificate Number: 2024M752760).

\bibliography{main.bib}

\appendix
\section{\large{End Matter}}
\twocolumngrid

\textit{Brief review of the symmetry-enriched Ising CFT.}---We briefly illustrate the physics of symmetry-enriched criticality~\cite{Verresen2021PRX} by considering the quantum phase transitions in the transverse-field Ising and cluster-Ising chain, both of which possess a global spin-flip $\mathbb{Z}_{2}$ symmetry and a spinless time-reversal symmetry $\mathbb{Z}_{2}^{T}$.
These transitions are described by the $(1+1)$D Ising CFT, featuring a unique local scaling operator $\sigma$ and a \emph{unique} nonlocal scaling operator $\mu$, both with scaling dimension $\Delta_{\sigma/\mu} = 1/8$~\cite{francesco2012conformal}.
At the lattice level, the local scaling operator $\sigma$ serves as the FM order parameter, while the nonlocal scaling operator $\mu$ corresponds to the Kramers–Wannier dual string order parameter of the symmetric phase—namely, the PM string operator in the transverse-field Ising model [see Eq.~\eqref{eq:opm}] and the SPT string operator in the cluster-Ising model [see Eq.~\eqref{eq:ospt}].
The nonlocal string operators can be interpreted as symmetry fluxes~\cite{Verresen2021PRX} that carry distinct time-reversal symmetry charge, $T \mu T = \pm \mu$, and exhibit the \emph{slowest} algebraic decay at both critical points, distinguishing two symmetry-enriched realizations of the same Ising CFT with distinct conformal boundary conditions~\cite{Yu2022PRL,Parker2018PRB}.
Importantly, these two realizations are topologically distinct and cannot be smoothly connected without encountering a multicritical point or explicitly breaking the protecting symmetries.

This picture also applies to the quasiperiodic setup studied in this work. It is noted that, physically, the distinct boundary criticalities exhibited by the trivial and topological QP-Ising critical points are not mere boundary artifacts of a single bulk fixed point. Instead, they are intrinsically rooted in the nontrivial bulk topological features of the quantum critical points, analogous to the bulk-boundary correspondence in gapped topological phases. 
Consequently, although their measured local bulk critical exponents agree within numerical accuracy, these critical points exhibit distinct symmetry-enriched realizations of the QP-Ising criticality.
To further substantiate this distinction, we map out the global phase diagram of a generalized quasiperiodic Ising Hamiltonian that interpolates between these two regimes and identify a multicritical point separating the two critical lines within the symmetry-preserving parameter space (see Sec.~II of the Supplemental Material for numerical results).

\textit{Physical observables.}---To characterize the rich phase diagram and critical behaviors of the model, we calculate various connected correlation functions and string order parameters. 
To detect the ferromagnetic (FM) order, we compute the spin-spin correlation
\begin{equation}
  \label{eq:cfm}
  C_\text{FM}(r) = \langle \sigma_{i}^{x} \sigma_{i+r}^{x} \rangle - \langle \sigma_{i}^{x} \rangle \langle \sigma_{i+r}^{x} \rangle \, .
\end{equation}
The trivial paramagnetic (PM) phase is characterized by the non-local PM string operator 
\begin{equation}
  \label{eq:opm}
  O_\text{PM}(r) = \langle \prod_{k=i}^{i+r-1} \sigma_{k}^{z} \rangle \, .
\end{equation}
To identify the symmetry-protected topological (SPT) cluster phase, we employ the SPT string order parameter
\begin{equation}
  \label{eq:ospt}
  O_\text{SPT}(r)  = \langle \sigma_{i}^{x} \sigma_{i+1}^{y} \left( \prod_{k=i+2}^{i+r-1} \sigma_{k}^{z} \right) \sigma_{i+r}^{y} \sigma_{i+r+1}^{x} \rangle \, .
\end{equation}
To complementarily characterize the nontrivial boundary critical behavior of the model, we consider the energy-energy correlation
\begin{equation}
  \label{eq:cen}
  C_\text{EN}(r) = \langle \epsilon_{i} \epsilon_{i+r}  \rangle - \langle \epsilon_{i} \rangle \langle \epsilon_{i+r} \rangle \, ,
\end{equation}
where the local energy operator is defined as $\epsilon_{i} \equiv \sigma_{i}^{x} \sigma_{i+1}^{x}$, which corresponds on the lattice to the thermal primary field in the language of conformal field theory (CFT)~\cite{francesco2012conformal,Yu2022PRL}.

On the other hand, quantum entanglement provides a powerful probe for quantum many-body systems.
We partition the chain of length $N$ into two equal parts $A = \{ 1, 2, \dots, N/2 \}$ and its complement.
The half-chain entanglement entropy is defined as $S_\text{vN} = - \text{Tr} [\rho_{A} \ln\rho_{A}]$ where $\rho_{A}$ is the reduced density matrix of the subsystem $A$ by tracing out its complement. 
For 1D critical states, $S_\text{vN}$ usually scales logarithmically with system size [periodic boundary condition is assumed here]: $S_\text{vN}(N) = \frac{c_\text{eff}}{3} \ln N + S_{0}$ where $c_\text{eff}$ is the effective central charge, which is a universal data for the criticality, and $S_{0}$ a non-universal constant.
To further reveal the topological features, we analyze the entanglement spectrum, defined as the set of the eigenvalues $\{ \xi_{n} \}$ of the entanglement Hamiltonian $\mathcal{H}_{A} \equiv - \ln \rho_{A}$. 
As indicated by the Li-Haldane conjecture~\cite{Li2008PRL,Yu2024PRL}, the low-lying part of $\{ \xi_{n} \}$ mimics the edge excitation spectrum of the physical Hamiltonian, serving as a fingerprint for topological edge degeneracies.

In the study of critical behaviors of the QP cluster-Ising model, proper averaging over the modulation phases $\phi_{1}$ and $\phi_{2}$ is essential. 
In the strong QP-modulation regime, correlation functions may oscillate in sign across different realizations. 
To prevent destructive interference during averaging, we compute the average of the absolute values of the correlations. 
The average over different samples is denoted by $\overline{\,\cdot\,}$ and for simplicity, the absolute value symbols $|\cdot|$ are implicitly assumed for correlations and string operators in the reported averaged results of this work.

\textit{Quasi-periodic SPT phase.}---In this section, we demonstrate that under strong QP-modulation, the cluster SPT phase is driven into an unconventional SPT phase that exhibits area-law entanglement scaling despite a vanishing bulk energy gap—a feature unique to QP systems, as we establish below.

\begin{figure}[tb]
  \centering
  \includegraphics[width=\linewidth]{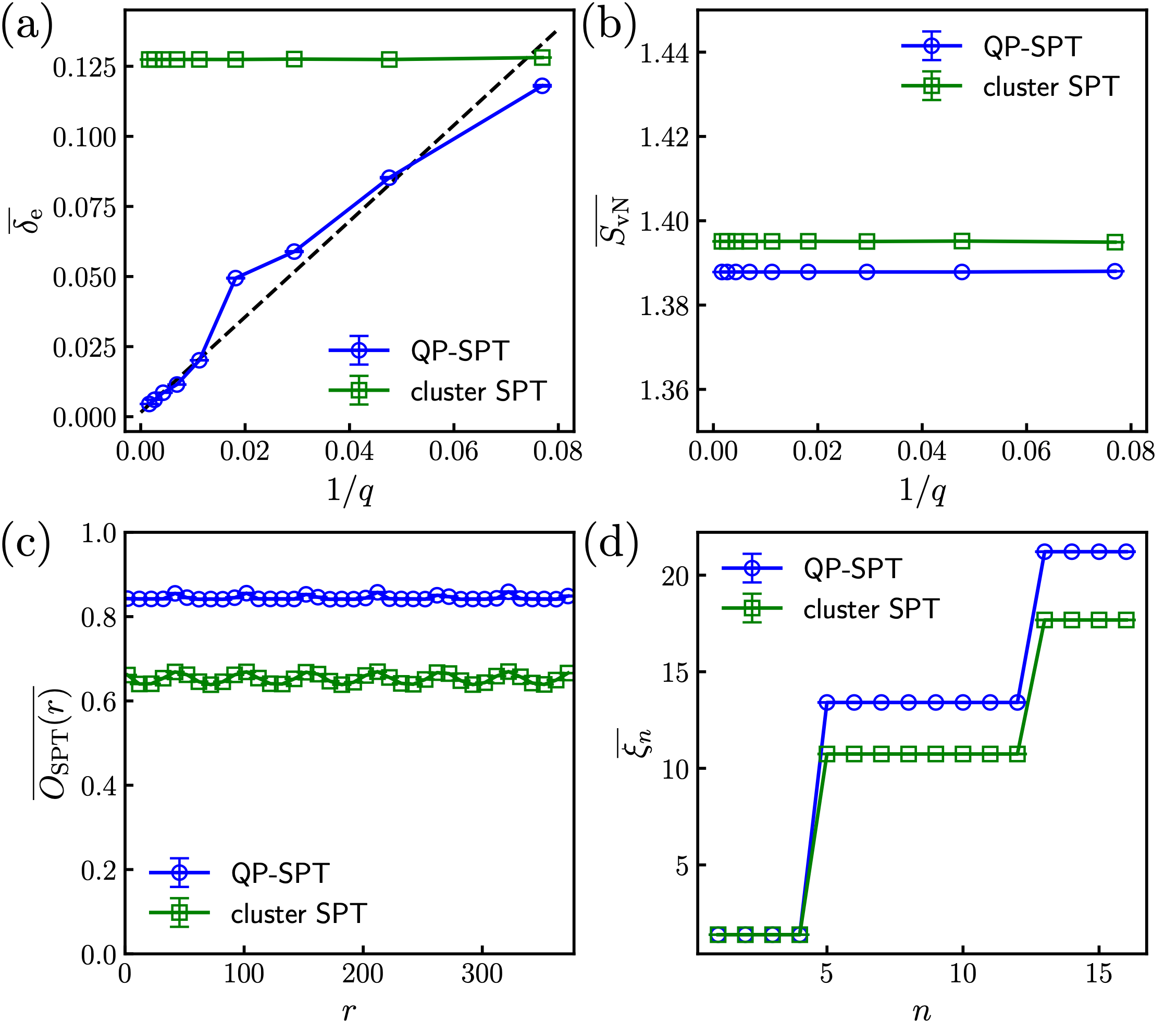}
  \caption{(a) Energy gap $\overline{\delta_\text{e}}$ versus $1/q$ for QP-SPT and cluster SPT phases. The black dashed line represents a fit to the scaling form $\overline{\delta_\text{e}}(q) \sim A/q^{\omega} + \overline{\delta_\text{e}}(q \to \infty)$, yielding $\omega \approx 1.0(2)$ with $\overline{\delta_\text{e}}(q \to \infty) \approx 0.002(2)$\,. (b) Area-law scaling of the half-chain  entanglement entropy $\overline{S_{\rm vN}}$. (c) Scaling behavior of the SPT string order parameter $\overline{O_{\rm SPT}(r)}$. (d) The bulk many-body entanglement spectrum $\overline{\xi_{n}}$. All  results are obtained under PBC. Parameters: $\bar{J} = \bar{g} = 1/2, h_{J}= 1, h_{g} = 2$ [the left-pointing triangle in Fig.~\ref{fig:phase_diagram}(c)] for QP-SPT and $\bar{J} = \bar{g} = 1/2, h_{J} = 1/2, h_{g} = 1/3$ [the right-pointing triangle in Fig.~\ref{fig:phase_diagram}(c)] for cluster SPT . The parameter $q$ ranges from $13$ to $610$ for (a,b) and is fixed at $q = 377$ and $610$, respectively, for (c) and (d).}
  \label{fig:phase}
\end{figure}

We focus on the phases by considering the representative point $\bar{J}=\bar{g}=1/2, h_J=1, h_g=2$, located in the lower-left region of the phase diagram Fig.~\ref{fig:phase_diagram}(c).
Finite-size scaling of the bulk energy gap [Fig.~\ref{fig:phase}(a)] and the half-chain entanglement entropy [Fig.~\ref{fig:phase}(b)] reveals that this phase is gapless yet obeys an area-law entanglement scaling, in sharp contrast to the conventional gapped cluster SPT phase.
At first glance, this behavior appears counterintuitive, as one-dimensional gapless systems with short-range interactions are generally expected to exhibit logarithmic entanglement scaling~\cite{Hastings_2007,Eisert2010RMP}.
We note that the $4$-fold degeneracy and the subsequent $8$-fold plateau reflect the combinatorial filling of the paired zero-energy and lowest-excited single-particle entanglement modes (see SM Sec.~IB for more details).
By further analyzing the SPT string order parameter [Fig.~\ref{fig:phase}(c)] and the entanglement spectrum [Fig.~\ref{fig:phase}(d)], we demonstrate the surprisingly stability of the SPT order with a gapless bulk, a regime we term the QP-SPT phase, which is unique to systems with QP-modulations.
Examining this stability is highly nontrivial and can not be easily captured by standard field theoretic methods since there is no disorder to average over, in contrast to disordered systems~\cite{Boyanovsky1982PRB,Kim1994PRB,Ye1993PRL}. Specifically, field-theoretical approaches for disordered systems~\cite{bernard1995perturbedconformalfieldtheory}—including the supersymmetric approach~\cite{MAASSARANI1997603} and the replica method with perturbative renormalization group and $\epsilon$-expansion~\cite{DOTSENKO2001445,Boyanovsky1982PRB,Kim1994PRB,Ye1993PRL}—rely on ensemble averaging over random disorder or calculations for short-range correlated randomness. In contrast, at the length scales $r<q_i$ of interest, the quasiperiodic systems studied here are deterministic and aperiodic, with long-range correlations. Consequently, no direct field-theoretical method has been developed for such deterministic structures. This fundamental distinction explains why standard approaches for disordered systems are not directly applicable and motivates our combination of exact solutions with numerical simulations.

To elucidate the underlying mechanism responsible for the emergence of the QP-SPT phase in QP systems, we note that the gapless nature of this regime originates from strong QP-modulations that effectively fragment the chain, rendering the system an assembly of weakly connected, gapped cluster-spin subchains [see Fig.~\ref{fig:phase_diagram}(b)].
Excitations localized near these weak links acquire arbitrarily low energies in the thermodynamic limit, i.e., $1/q \to 0$, leading to an overall gapless spectrum despite each individual subchain being gapped.
Consequently, the gapless excitations are intrinsic to QP-modulation and are not associated with rare-region effects, in sharp contrast to the Griffiths–McCoy phase in disordered spin chains~\cite{McCoy1968PR,McCoy1969PR,Griffiths1969PRL}.
By contrast, the entanglement structure and correlation properties are dominated by the universal behavior of each gapped subchain and are largely insensitive to the weak couplings.
As a result, the system exhibits area-law entanglement scaling and long-range order in SPT string order parameters, analogous to conventional gapped SPT phases.

\textit{Comparison to the infinite-quasiperiodicity fixed point.}---It is noted that an infinite-quasiperiodicity fixed point has been found in Refs.~\cite{Agrawal2020NC,Agrawal2020PRL} for Heisenberg and $q$-state Potts models. Crucially, the absence of an infinite-quasiperiodicity fixed point in our model is a distinct physical consequence of the Ising-like model, rather than a limitation of the strong-disorder renormalization group (SDRG) approach. In contrast to the Heisenberg or $q$-state Potts chains where strong quasiperiodic modulations preserve the quasiperiodic sequence structure under RG flows—leading to an infinite-quasiperiodicity fixed point with an activated dynamical scaling $\overline{\delta_\text{e}} \sim \text{e}^{-a\ln^2 L}$ and an infinite dynamical exponent $z=\infty$—the quasiperiodic sequence structure in the Ising case is not preserved. Consequently, the low-energy critical behavior remains governed by a quasiperiodic Ising fixed point featuring a finite dynamical exponent $z \approx 1.9$ and a power-law energy gap $\overline{\delta_\text{e}} \sim L^{-z}$ as reported in this work and previous literature~\cite{Crowley2018PRL,Crowley_2022}.

\end{document}


\let\oldaddcontentsline\addcontentsline
\renewcommand{\addcontentsline}[3]{}
\let\addcontentsline\oldaddcontentsline
\onecolumngrid

\clearpage
\newpage
\widetext

\begin{center}
\textbf{\large Supplemental Material for ``Topological Quantum Criticality in Quasiperiodic Cluster Ising Chain''}
\end{center}

\author{Sheng Yang}
\affiliation{Institute for Advanced Study in Physics and School of Physics, Zhejiang University, Hangzhou 310058, China}

\author{Hai-Qing Lin}
\email{hqlin@zju.edu.cn}
\affiliation{Institute for Advanced Study in Physics and School of Physics, Zhejiang University, Hangzhou 310058, China}

\author{Xue-Jia Yu}
\email{xuejiayu@eitech.edu.cn}
\affiliation{Eastern Institute of Technology, Ningbo 315200, China}

\maketitle

\renewcommand{\thefigure}{S\arabic{figure}}
\setcounter{figure}{0}
\renewcommand{\theequation}{S\arabic{equation}}
\setcounter{equation}{0}
\renewcommand{\thesection}{\Roman{section}}
\setcounter{section}{0}
\setcounter{secnumdepth}{4}

\addtocontents{toc}{\protect\setcounter{tocdepth}{0}}
{
\tableofcontents
}

\section{Details of the analytical calculations for the quasi-periodic cluster-Ising chain}
\label{S1}

\subsection{Diagonalization of the Hamiltonian}

The quasi-periodic (QP) modulated cluster-Ising (CI) model is described by the Hamiltonian [periodic boundary condition (PBC) is assumed here]:
\begin{equation}
    H_\text{CI}^\text{QP} = - \sum_{i=1}^{N-1} J_{i} \sigma_{i}^{x} \sigma_{i+1}^{x} - \sum_{i=1}^{N-2} g_{i} \sigma_{i}^{x} \sigma_{i+1}^{z} \sigma_{i+2}^{x} - \left( g_{N-1} \sigma_{N-1}^{x} \sigma_{N}^{z} \sigma_{1}^{x} + g_{N} \sigma_{N}^{x} \sigma_{1}^{z} \sigma_{2}^{x} + J_{N} \sigma_{N}^{x} \sigma_{1}^{x} \right) \, ,
\end{equation}
where $J_{i} = \bar{J} + h_{J} \cos{[Q(i+1/2)+\phi_{1}]}$ and $g_{i} = \bar{g} + h_{g} \cos{[Qi+\phi_{1}+\phi_{2}]}$ with $Q/(2\pi) = \tau_\text{G}$ and $\tau_\text{G} = (\sqrt{5} + 1 ) / 2$ the Golden ratio.

To solve this Hamiltonian, we perform the Jordan-Wigner transformation, 
\begin{equation}
    \label{eq:Hspin}
    \sigma_{j}^{x} = \prod_{i<j} \left( 1-2c_{i}^{\dagger}c_{i}^{} \right) (c_{j}^{\dagger} + c_{j}^{}) \, , \qquad \sigma_{j}^{y} = \text{i} \prod_{i<j} \left( 1-2c_{i}^{\dagger}c_{i}^{} \right) (c_{j}^{\dagger} - c_{j}^{}) \, , \qquad \sigma_{j}^{z} = 1 - 2 c_{j}^{\dagger} c_{j}^{} \, ,
\end{equation}
after which the spin Hamiltonian can be recast into the form:
\begin{align}
    \label{eq:Hfermion}
    H = & - \sum_{i=1}^{N-1} J_{i} \left( c_{i}^{\dagger}c_{i+1}^{\dagger} + c_{i}^{\dagger}c_{i+1}^{} + c_{i+1}^{\dagger}c_{i}^{} + c_{i+1}^{}c_{i}^{} \right) - \sum_{i=1}^{N-2} g_{i} \left( c_{i}^{\dagger}c_{i+2}^{\dagger} + c_{i}^{\dagger}c_{i+2}^{} + c_{i+2}^{\dagger}c_{i}^{} + c_{i+2}^{}c_{i}^{} \right) \nonumber \\
    & + \text{e}^{\text{i}\pi\sum_{j=1}^{N}c_{j}^{\dagger}c_{j}^{}} J_{N} \left( c_{N}^{\dagger}c_{1}^{\dagger} + c_{N}^{\dagger}c_{1}^{} + c_{1}^{\dagger}c_{N}^{} + c_{1}^{}c_{N}^{} \right) \nonumber \\
    & + \text{e}^{\text{i}\pi\sum_{j=1}^{N}c_{j}^{\dagger}c_{j}^{}} g_{N-1} \left( c_{N-1}^{\dagger}c_{1}^{\dagger} + c_{N-1}^{\dagger}c_{1}^{} + c_{1}^{\dagger}c_{N-1}^{} + c_{1}^{}c_{N-1}^{} \right) \nonumber \\ 
    & + \text{e}^{\text{i}\pi\sum_{j=1}^{N}c_{j}^{\dagger}c_{j}^{}} g_{N} \left( c_{N}^{\dagger}c_{2}^{\dagger} + c_{N}^{\dagger}c_{2}^{} + c_{2}^{\dagger}c_{N}^{} + c_{2}^{}c_{N}^{} \right) \, .
\end{align}
The system decomposes into two disjoint sectors characterized by the parity of the total fermion-number operator $n_\text{tot} = \sum_{i=1}^{N} c_{i}^{\dagger}c_{i}^{}$. 
More specifically, the even-parity sector corresponds to a free-fermion Hamiltonian with anti-PBC whereas the odd-parity sector obeys PBC. 
To determine the true ground state of the original spin model, one must compare the lowest energy eigenvalues obtained from both sectors.
We can anticipate the global ground state to be the vacuum state (with $n_\text{tot} = 0$) within the anti-PBC  sector, which is also confirmed in our numerical calculations. 
For excited states, parity constraints imply that the relevant low-energy candidates are the $n_\text{tot} = 1$ state in the PBC sector and the $n_\text{tot} = 2$ state in the anti-PBC sector. 
The energy gap $\delta_\text{e}$ is thus determined by the difference between the energy of the global ground state and the lower of these two excited energy levels.

Using the Nambu formalism, we define the spinor vector $\Psi^{\dagger} \equiv \begin{pmatrix} \mathbf{c}^{\dagger} & \mathbf{c}^{} \end{pmatrix} = \begin{pmatrix} c_{1}^{\dagger} & \dots & c_{N}^{\dagger} & c_{1}^{} & \dots & c_{N}^{} \end{pmatrix}$ and its conjugation $\Psi$, each of length $2N$.
The free-fermion Hamiltonian~\eqref{eq:Hfermion} can be written in the standard Bogoliubov-de Gennes (BdG) form in terms of $\Psi$ as
\begin{equation}
    H = \frac{1}{2} \Psi^{\dagger} \mathcal{H}_\text{BdG} \Psi + \text{const.} \qquad \mathcal{H}_\text{BdG} = \begin{pmatrix} \text{A} & \text{B} \\ -\text{B} & -\text{A} \end{pmatrix} \, ,
\end{equation}
where $\text{A}$ ($\text{B}$) is a real symmetric (anti-symmetric) matrix. 
To diagonalize the Hamiltonian~\eqref{eq:Hfermion}, we introduce a unitary transformation $U$, whose columns are eigenvectors of $\mathcal{H}_\text{BdG}$. 
This matrix defines the transformation to the Bogoliubov quasiparticle operators
\begin{equation}
    \begin{pmatrix}
        \bm{\eta} \\ \bm{\eta}^{\dagger}
    \end{pmatrix}
    = U^{\dagger} \Psi = U^{\dagger}
    \begin{pmatrix}
        \mathbf{c} \\ \mathbf{c}^{\dagger}
    \end{pmatrix} \, .
\end{equation}
Consequently, the Hamiltonian is diagonalized as
\begin{equation}
    H = \frac{1}{2} \begin{pmatrix} \bm{\eta}^{\dagger} & \bm{\eta}^{} \end{pmatrix} \left( U^{\dagger} \mathcal{H}_\text{BdG} U^{} \right) \begin{pmatrix} \bm{\eta}^{} \\ \bm{\eta}^{\dagger} \end{pmatrix} = \sum_{k=1}^{N} \epsilon_{k} \eta_{k}^{\dagger} \eta_{k}^{} + \text{const.} \, ,
\end{equation}
where $\epsilon_{k} \geq 0$ are the positive eigenvalues of $\mathcal{H}_\text{BdG}$. 
It is noted that eigenvalues appear in pairs $\pm \epsilon_{k}$ in $\mathcal{H}_\text{BdG}$ due to the particle-hole symmetry.
The ground state $|\text{GS}\rangle$ is then the vacuum of these quasiparticles, satisfying $\eta_{k} |\text{GS}\rangle = 0$ for all $k = 1$ to $N$.

Explicitly, the quasiparticle operators $\eta_{k}^{}$ are related to the original fermions $c_{i}^{}$ via the coefficients of $U$
\begin{equation}
    c_{i}^{} = \sum_{k=1}^{N} U_{i,k} \eta_{k}^{} + \sum_{k=1}^{N} U_{i,k+N} \eta_{k}^{\dagger} \quad \text{ and } \quad c_{i}^{\dagger} = \sum_{k=1}^{N} U_{i,k} \eta_{k}^{\dagger} + \sum_{k=1}^{N} U_{i,k+N} \eta_{k}^{} \, ,
\end{equation}
where we have used the fact that $U$ can be chosen to be real.
Then the two-point correlations can be evaluated using the   condition $\eta_{k}^{} |\text{GS}\rangle = 0$. 
For example, 
\begin{equation}
    \langle c_{i}^{\dagger} c_{j}^{} \rangle = \sum_{kk'} \langle \text{GS} | ( U_{i,k} \eta_{k}^{\dagger} + U_{i,k+N} \eta_{k}^{} ) ( U_{j,k'} \eta_{k'}^{} + U_{j,k'+N} \eta_{k'}^{\dagger} ) | \text{GS} \rangle = \sum_{k=1}^{N} U_{i,k+N} U_{j,k+N} = \sum_{k=1}^{N} U_{i,k+N}^{} U^{\dagger}_{k+N,j} \, .
\end{equation}
Similarly, we have
\begin{equation}
    \langle c_{i}^{\dagger} c_{j}^{\dagger} \rangle = \sum_{k=1}^{N} U_{i,k+N}^{} U_{k,j}^{\dagger} \, , \quad{} \langle c_{i}^{} c_{j}^{\dagger} \rangle = \sum_{k=1}^{N} U_{i,k}^{} U_{k,j}^{\dagger} \, , \quad \text{and} \quad \langle c_{i}^{} c_{j}^{} \rangle = \sum_{k=1}^{N} U_{i,k}^{} U_{k+N,j}^{\dagger} \, .
\end{equation}
Combining these results, the full correlation matrix $\mathcal{G} \equiv \langle \text{GS} | \Psi \Psi^{\dagger} | \text{GS} \rangle$ is obtained. 
For convenience's sake, we define $G_{ij} \equiv \langle c_{i}^{} c_{j}^{\dagger} \rangle$ the normal correlation matrix and $F_{ij} \equiv \langle c_{i}^{} c_{j}^{} \rangle$ the anomalous correlation matrix, respectively.
Note that the former is Hermitian $G = G^{\dagger}$, while the latter is antisymmetric $F^\text{T} = - F$ (due to the fermionic anti-commutations).
These correlation matrices effectively capture all information of the Gaussian ground state and serve as the starting point for computing the quantities investigated in this work~\cite{Glen2024SciPost}.

\subsection{Von Neumann entanglement entropy and many-body entanglement spectrum}

First, the Von Neumann entanglement entropy $S_\text{vN}$ of a contiguous interval of $l$ sites (denoted by $\mathcal{A}_{l}$) is determined by the reduced density matrix $\rho_{l} = \text{Tr}_{N-l} | \text{GS} \rangle \langle \text{GS} |$, which is a Gaussian state. 
This means that $\rho_{l}$ is fully characterized by the block correlation matrix $\mathcal{G}|_{\mathcal{A}_l}$, i.e., $\mathcal{G}$ restricted to region $\mathcal{A}_{l}$.
By diagonalizing the block correlation matrix $\mathcal{G}|_{\mathcal{A}_{l}}$, we obtain its eigenvalues $ \{\varphi_{n} \}_{n=1}^{2l}$, which is often called the single-particle entanglement spectrum in the literature.
It is noted that, due to the particle-hole symmetry, $\{ \varphi_{n} \}_{n=1}^{2l}$ again appear in pairs in the form $\{ \varphi_{n}, 1-\varphi_{n} \}_{n=1}^{l}$ that is symmetric around $1/2$. 
The entanglement entropy can be computed accordingly by~\cite{Chung2001prb,Cheong2004prb,Peschel2009jpa}
\begin{equation}
    S_\text{vN} = - \sum_{n=1}^{l} \left[ \varphi_{n} \ln \varphi_{n} + (1 - \varphi_{n}) \ln (1 - \varphi_{n}) \right] \, ,
\end{equation}
where $\varphi_{n}$ with $n=1,\cdots,l$ are $l$ representative eigenvalues chosen from each particle-hole symmetric pair.
In addition, we can define the entanglement energies as~\cite{Cheong2004prb}
\begin{equation}
    \tilde{\epsilon}_{n} = \ln\left( \frac{1-\varphi_{n}}{\varphi_{n}} \right) \, ,
\end{equation}
which is obviously symmetric around $0$ in the form $\{ \tilde{\epsilon}_{n}, - \tilde{\epsilon}_{n} \}_{n=1}^{l}$. 
To construct the physical many-body entanglement spectrum, we restrict ourselves to the positive branch. 
The many-body entanglement spectrum can be determined by filling these physical modes. 
In particular, the entanglement Hamiltonian $H_{\rm E} = - \ln{\rho_{l}}$ is diagonal in the eigen-basis of the block correlation matrix:
\begin{equation}
    H_{\rm E} = E_{\rm shift} + \sum_{n=1}^{l} \tilde{\epsilon}_{n} \alpha_{n}^{\dagger} \alpha_{n}^{} \, ,
\end{equation}
where $\alpha_{n}^{\dagger}$ creates an excitation mode with energy $\tilde{\epsilon}_{n} \geq 0$.

The shifted energy can be determined by the normalization condition ${\rm Tr} \rho_{l} = 1$ and
\begin{equation}
    E_{\rm shift} = \ln \prod_{n=1}^{l} (1 + \text{e}^{-\tilde{\epsilon}_{n}}) = \sum_{n=1}^{l} \ln \left( 1 + \text{e}^{-\tilde{\epsilon}_{n}} \right) \,.
\end{equation}
Therefore, each eigenvalue of the entanglement Hamiltonian $H_\text{E}$ corresponds to a specific occupation configuration $\mathbf{m} = (m_{1}, m_{2}, \dots, m_{l})$ where $m_{i} \in \{ 0, 1 \}$,
\begin{equation}
    \xi_{\mathbf{m}} = \sum_{n=1}^{l} m_{n} \tilde{\epsilon}_{n} + \sum_{n=1}^{l} \ln \left( 1 + \text{e}^{-\tilde{\epsilon}_{n}} \right) \, .
\end{equation}
In this work, we focus on the low-lying part of the spectrum, which can be obtained efficiently by using the min-heap algorithm. 

Based on this construction, the specific structure seen in Fig.~4(d) in the main text can be understood as follows:
\begin{itemize}
    \item[1.] \emph{The four-fold degeneracy ($n = 1$ to $4$)}: The $\mathbb{Z}_{2} \times \mathbb{Z}_{2}^{T}$ symmetry protects two localized topological zero modes at the entanglement cuts, corresponding to two zero-energy single-particle levels, say $\tilde{\epsilon}_{1}$ and $\tilde{\epsilon}_{2}$. The $2^{2} = 4$ possible occupancy configurations of these two zero modes generate the four-fold degenerate lowest-lying states of the many-body entanglement spectrum; 
    \item[2.] \emph{The jump at $n = 5$ and the following eight-fold degenerate plateau ($n = 5$ to $12$)}: The first jump corresponds to exciting the lowest non-zero single-particle levels. In our model, due to the same protecting symmetry, these excitations also appear in nearly degenerate pairs, say $\tilde{\epsilon}_{3} \approx \tilde{\epsilon}_{4} > 0$. Exciting only \emph{one} of these two modes gives $2$ configurations. Multiplying this by the four-fold zero-energy degeneracy yields $2\times4=8$ many-body states with finite entanglement energy. This perfectly accounts for the plateau from  $n = 5$ to $12$;
    \item[3.] \emph{The jump at $n=13$}: This second jump occurs when both $\tilde{\epsilon}_{3}$ and $\tilde{\epsilon}_{4}$ are simultaneously occupied or higher-energy single-particle modes begin to be occupied.
\end{itemize}

\subsection{Correlation functions and string order parameters}

To compute the correlation functions $C_\text{FM}(r)$ and $C_\text{EN}(r)$, as well as nonlocal string operators $O_\text{PM}(r)$ and $O_\text{SPT}(r)$, it is convenient to use the language of Majorana operators $\gamma_{2i-1} = c_{i}^{\dagger} + c_{i}^{} = A_{i}$ and $\gamma_{2i} = \text{i} ( c_{i}^{\dagger} - c_{i}^{} ) = \text{i} B_{i}$. 
Noticing that $A_{i}B_{i} = 1 - 2 c_{i}^{\dagger} c_{i}^{}$, these observables can be expressed as strings of $A_{i}$ and $B_{i}$~\cite{Glen2024SciPost}. 
For example, 
\begin{equation}
    \label{eq:cfm_string}
    \langle \sigma_{i}^{x} \sigma_{j}^{x} \rangle = \langle B_{i} (A_{i+1} B_{i+1}) \cdots (A_{j-1} B_{j-1}) A_{j} \rangle \, ,
\end{equation}
where we have assumed $i < j$. 
Since the ground state is a Gaussian state, according to Wick's theorem, the expectation value of any string of Majorana operators can be expressed in terms of two-point Majorana correlations. 
Furthermore, based on the observation that 
\begin{align}
    \langle A_{i} A_{j} \rangle & = \langle (c_{i}^{\dagger} + c_{i}^{}) (c_{j}^{\dagger} + c_{j}^{}) \rangle = \delta_{ij} + F_{ij} + F_{ji}^{*} = \delta_{ij} \, , \\ 
    \langle B_{i} B_{j} \rangle & = \langle (c_{i}^{\dagger} - c_{i}^{}) (c_{j}^{\dagger} - c_{j}^{}) \rangle = - \delta_{ij} + F_{ij} + F_{ji}^{*} = - \delta_{ij} \, ,
\end{align}
we only need to consider two-point correlations between the $A$-type and $B$-type operators in Eq.~\eqref{eq:cfm_string}.
Finally, the expectation becomes
\begin{equation}
    \langle \sigma_{i}^{x} \sigma_{j}^{x} \rangle = \det \begin{bmatrix} M_{i,i+1} & M_{i,i+2} & \cdots & M_{i,j} \\ M_{i+1,i+1} & M_{i+1,i+2} & \cdots & M_{i+1,j} \\ \vdots & \vdots & \ddots & \vdots \\ M_{j-1,i+1} & M_{j-1,i+2} & \cdots & M_{j-1,j} \end{bmatrix} \, ,
\end{equation}
where we have defined $M_{k,l} \equiv \langle B_{k} A_{l} \rangle = \delta_{kl} - 2 ( G_{kl} + F_{kl} )$.
Similarly, we can calculate the energy-energy correlation
\begin{equation}
    C_\text{EN}(i,j) = \langle \sigma_{i}^{x} \sigma_{i+1}^{x} \sigma_{j}^{x} \sigma_{j+1}^{x} \rangle - \langle \sigma_{i}^{x} \sigma_{i+1}^{x} \rangle \langle \sigma_{j}^{x} \sigma_{j+1}^{x} \rangle = \langle B_{i} A_{i+1} B_{j} A_{j+1} \rangle - \langle B_{i} A_{i+1} \rangle \langle B_{j} A_{j+1} \rangle = - M_{i,j+1} M_{j,i+1} \, ,
\end{equation}
the PM string order parameter
\begin{equation}
    O_\text{PM}(i,j) = \langle \prod_{k=i}^{j} \sigma_{k}^{z} \rangle = \langle (A_{i} B_{i}) \cdots (A_{j} B_{j}) \rangle = (-1)^{j-i+1} \det \begin{bmatrix} M_{i,i} & M_{i,i+1} & \cdots & M_{i,j} \\ M_{i+1,i} & M_{i+1,i+1} & \cdots & M_{i+1,j} \\ \vdots & \vdots & \ddots & \vdots \\ M_{j,i} & M_{j,i+1} & \cdots & M_{j,j} \end{bmatrix} \, ,
\end{equation}
and the SPT string order parameter
\begin{equation}
    O_\text{SPT}(i,j) = \langle \sigma_{i}^{x} \sigma_{i+1}^{y} \left( \prod_{k=i+2}^{j-1} \sigma_{k}^{z} \right) \sigma_{j}^{y} \sigma_{j+1}^{x} \rangle = (-1)^{j-i} \det \begin{bmatrix} M_{i,i+2} & M_{i,i+3} & \cdots & M_{i,j+1} \\ M_{i+1,i+2} & M_{i+1,i+3} & \cdots & M_{i+1,j+1} \\ \vdots & \vdots & \ddots & \vdots \\ M_{j-1,i+2} & M_{j-1,i+3} & \cdots & M_{j-1,j+1} \end{bmatrix} \, .
\end{equation}
Using these expressions, we can compute $C_\text{FM}(r)$, $C_\text{EN}(r)$, $O_\text{PM}(r)$, and $O_\text{SPT}(r)$ for each realization of $\phi_{1}$ and $\phi_{2}$. 
For PBC, the average of the absolute value of these quantities is performed on different samples of $(\phi_{1}, \phi_{2})$ and on different sites of the system by fixing the distance $r$ within each sampling realization.
The sampling number is $10^4$ for $q < 100$ and $10^3$ for $q > 100$ in our work.
For OBC, the average of boundary-bulk correlations is performed only on different samples with the sampling number $10^{5}$. 
For the averaged gap $\overline{\delta_\text{e}}$, the entanglement entropy $\overline{S_\text{vN}}$, and the many-body entanglement spectrum $\{ \overline{\xi_{n}} \}$, the sampling number is $10^5$ for all $q$ under PBC.

Finally, we note that the ground state of the QP cluster-Ising chain becomes two-fold degenerate under OBC due to the spontaneous magnetization on the edge. 
In this case, the boundary-bulk spin-spin correlation is calculated for the second site with the bulk instead, i.e., $C_\text{FM}(2,2+r)$, in the same way as in Ref.~\cite{Yu2022PRL}.

\subsection{The phase boundaries}

In fact, the phase boundaries can be exactly determined by analyzing the localization length of the Majorana modes.
By mapping to Majorana representation, the spin Hamiltonian under OBC can be written as:
\begin{equation}
    H = - \sum_{i=1}^{N-1} J_{i} \sigma_{i}^{x} \sigma_{i+1}^{x} - \sum_{i=1}^{N-2} g_{i} \sigma_{i}^{x} \sigma_{i+1}^{z} \sigma_{i+2}^{x} = \text{i} \sum_{i=1}^{N-1} J_{i} \gamma_{2i} \gamma_{2i+1} + \text{i} \sum_{i=1}^{N-2} g_{i} \gamma_{2i} \gamma_{2i+3} \, .
\end{equation}
To determine the zero-energy modes, we start with the anzatz $\zeta = \sum_{k=1}^{2N} \psi_{k} \gamma_{k}$ by solving the commutation relation $[H, \zeta] = 0$. 
Due to the specific coupling structure, there exist decoupled modes at the boundaries; 
specifically, $\zeta_\text{L} \propto \gamma_{1}$ and $\zeta_\text{R} \propto \gamma_{2N}$ are exact zero modes with $\psi_{k} = 0$ for $k \neq 1$ and $k \neq 2N$, respectively.

We can also identify two nontrivial solutions $\zeta_\text{L/R} = \sum_{k=1}^{2N} \psi_{k}^\text{L/R} \gamma_{k}$, which is governed by the recurrence relation
\begin{align}
    \psi_{2i+3}^\text{L} = -\frac{J_{i}}{g_{i}} \psi_{2i+1}^\text{L} = \psi_{3}^\text{L} \prod_{j=1}^{i} \left( - \frac{J_{j}}{g_{j}} \right) \, , \\
    \psi_{2i-2}^\text{R} = -\frac{J_{i}}{g_{i-1}} \psi_{2i}^\text{R} = \psi_{2N-2}^\text{R} \prod_{j=i}^{N-1} \left( -\frac{J_{j}}{g_{j-1}} \right) \, .
\end{align}
Focusing on the left-localized mode, the amplitude of the wavefunction behaves as 
\begin{equation}
    \left| \psi_{2i+3}^\text{L} \right| \propto \prod_{k \leq i} \left| \frac{J_{k}}{g_{k}} \right| \equiv \text{exp} \left( - \sum_{k \leq i} \delta_{i} \right)
\end{equation}
where $\delta_{i} \equiv \ln(|g_{i}/J_{i}|)$ is the local reduced coupling. 
Therefore, the localization length of this edge mode can be defined by $\left| \psi_{2l+3}^\text{L} \right| \sim \exp(-l/l_{0})$ with $1/l_{0} = \frac{1}{l} \sum_{i=1}^{l} \delta_{i}$. 
It is clear that the localization length $l_{0}$ diverges when $[\delta_{i}] = 0$ in the limit of $l \to \infty$ ($[ \cdot ]$ means the average over all sites $i$). 
It also serves as the condition for the criticality and can be used to identify the phase boundaries.
Setting $\bar{J} = \bar{g}$ (as in the main text), we can obtain three solutions. 
The condition $h_{J} = h_{g}$ represents the vertical boundary in Fig.~1(c) in the main text, while the remaining two boundaries [the dashed lines in Fig.~1(c) in the main text] are described, respectively, by~\cite{Crowley_2022}
\begin{align}
    \frac{\bar{J}}{h_{g}} & = \frac{1 + (h_{J}/h_{g})^{2}}{2} \quad{} \text{for } h_{J} < h_{g} \, , \\
    \frac{\bar{J}}{h_{J}} & = \frac{1 + (h_{g}/h_{J})^{2}}{2} \quad{} \text{for } h_{J} > h_{g} \, .
\end{align}
It is clear that the phase diagram should be symmetric under the interchange of $h_{J}$ and $h_{g}$.
The precisely identification of the phase boundaries paves the way for our analyses of the criticalities.

\section{Additional numerical results for the transition between the QP-FM and QP-SPT phases}
\label{S2}

In this section, we provide additional results for the QP-Ising and QP-Topological Ising transitions.

\begin{figure}[t]
  \centering
  \includegraphics[width=0.25\linewidth]{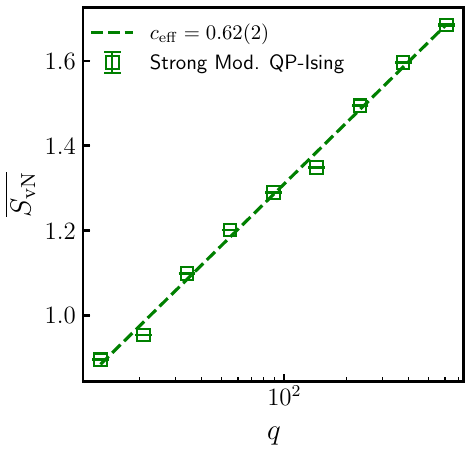}
  \caption{Logarithmic-law behavior of the half-chain entanglement entropy $\overline{S_{\rm vN}} = \frac{c_\text{eff}}{3} \ln q + {\rm const}$ at the critical point $\bar{J} = \bar{g} = 1/2, h_{J} = h_{g} = 1$ for the strongly modulated QP-Ising chain. The parameter $q$ ranges from $13$ to $610$ under PBC.}
  \label{fig:trivial_ceff}
\end{figure}

First, we examine the logarithmic scaling behavior of the half-chain entanglement entropy $\overline{S_\text{vN}} \sim \frac{c_\text{eff}}{3} \ln q + \text{const}$ at a representative critical point of the strongly modulated QP-Ising chain (e.g., $\bar{J} = \bar{g} = 1/2$ and $h_{J} = h_{g} = 1$). 
As shown in Fig.~\ref{fig:trivial_ceff}, the extracted effective central charge $c_\text{eff} = 0.62(2)$ is in remarkable agreement with the value $c_\text{eff} = 0.63(2)$ reported in the main text for the QP-Topological Ising model with the same parameter setting. 
This result further strengthens our claim of the bulk indistinguishability between the QP-Ising and QP-Topological Ising transitions (see the main text for details).

\begin{figure}[t]
  \centering
  \includegraphics[width=0.5\linewidth]{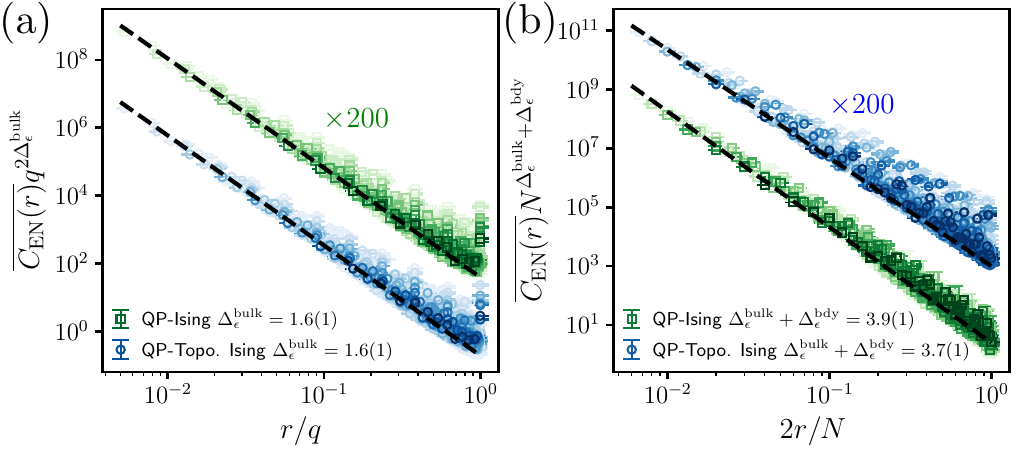}
  \caption{Data collapse of (a) bulk-bulk and (b) boundary-bulk energy-energy correlation functions $\overline{C_\text{EN}(r)}$ for QP-Ising and QP-Topological Ising transitions. The dashed lines are $\propto 1/(r/q)^{2\Delta_{\epsilon}^\text{bulk}}$ or $\propto 1/(2r/N)^{\Delta_{\epsilon}^\text{bulk} + \Delta_{\epsilon}^\text{bdy}}$ for comparison with the data collapse. To achieve a better visualization, we have multiplied an additional global factor for the correlations of QP-Ising or QP-Topological Ising as indicated by ``$\times 200$''. Parameters: $\bar{J} = \bar{g} = 1/2, h_{J} = h_{g} = 1$ [the red star in Fig.~1(c) in the main text]. The parameter $q$ ranges from $13$ to $377$ in (a) under PBC, and $N$ ranges from $100$ to $600$ in (b) under OBC.}
  \label{fig:energy_correlation}
\end{figure}

Second, the scaling analysis of the bulk-bulk and boundary-bulk energy-energy correlations is performed in Fig.~\ref{fig:energy_correlation}. 
It shows that the energy-energy correlation has the same bulk and boundary exponents, i.e., $\Delta_{\epsilon}^\text{bulk}$ and $\Delta_{\epsilon}^\text{bdy}$, for QP-Ising and QP-Topological Ising transitions. 
This observation is similar to the clean case, in which we have $\Delta_{\epsilon}^\text{bulk} = 1$ and $\Delta_{\epsilon}^\text{bdy} = 2$ for both clean Ising and clean topological Ising transitions~\cite{Yu2022PRL}.

\begin{figure}[t]
  \centering
  \includegraphics[width=\linewidth]{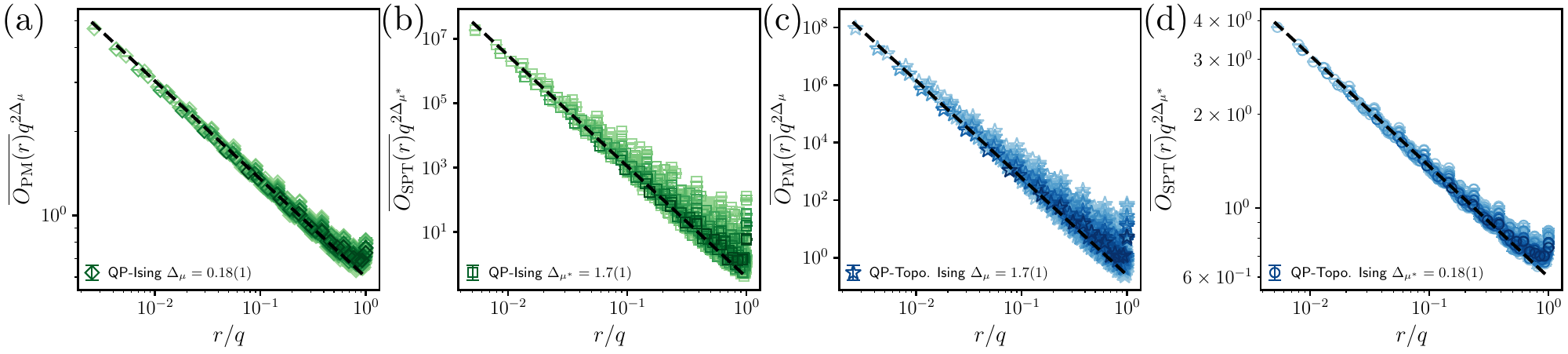}
  \caption{Data collapse of string operators $\overline{O_\text{PM/SPT}(r)}$ for QP-Ising and QP-Topological Ising transitions, respectively. The black dashed lines are $\propto 1/(r/q)^{2\Delta_{\mu/\mu^*}}$ for comparison with the data collapse. Parameters: $\bar{J} = \bar{g} = 1/2, h_{J} = h_{g} = 1$ [the red star in Fig.~1(c) in the main text]. The parameter $q$ ranges from $13$ to $377$ under PBC.}
  \label{fig:string}
\end{figure}

Third, we calculate the averaged nonlocal string order parameters $\overline{O_\text{PM}(r)}$ and $\overline{O_\text{SPT}(r)}$ for both QP-Ising and QP-Topological Ising transitions as shown in Fig.~\ref{fig:string}. 
We find that the decaying behaviors of $\overline{O_\text{PM}(r)}$ and $\overline{O_\text{SPT}(r)}$ are totally distinct at QP-Ising and QP-Topological Ising transitions. 
In particular, for the QP-Ising transition, the PM string order $\overline{O_\text{PM}(r)}$ decays much more slowly than the SPT string order $\overline{O_\text{SPT}(r)}$, while the situation reverses for the QP-Topological Ising transition;
this can be understood by the fact that the QP transverse-field Ising and QP cluster-Ising models can be mapped onto each other with a unitary transformation, which effectively interchanges the PM and SPT string order parameters.
It is noted that a slower decay of the SPT string order relative to the PM string order serves as a characteristic signature of the nontrivial topology of the transition, which is also supported by the two-fold degeneracy of the many-body entanglement spectrum shown in Fig.~3(d) in the main text. 

\begin{figure}[t]
  \centering
  \includegraphics[width=0.75\linewidth]{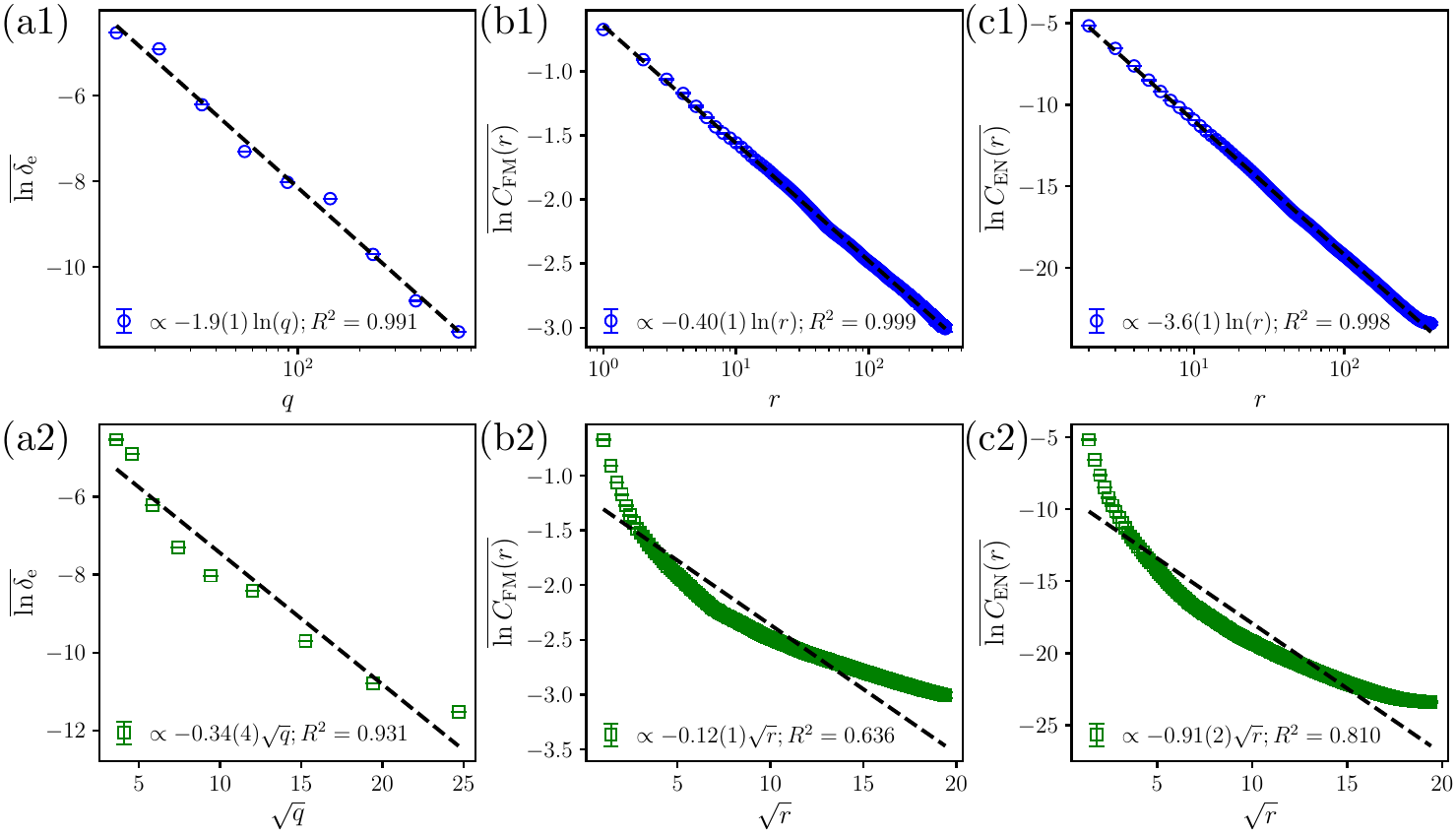}
  \caption{Scaling analysis of typical energy gap $\overline{\ln\delta_\text{e}}$, the typical spin-spin correlation $\overline{\ln C_\text{FM}(r)}$, and the typical energy-energy correlation $\overline{\ln C_\text{EN}(r)}$ at $\bar{J} = \bar{g} = 1/2$ and $h_{J} = h_{g} = 1$ [the red star in Fig.~1(c) in the main text] for the QP cluster-Ising model. The parameter $q$ ranges from $13$ to $610$ for (a1-a2), and $q = 377$ for (b1-b2) and (c1-c2) under PBC. Note the logarithmic and linear scales of the horizontal axes in the upper and lower panels, respectively. The goodness of fit is quantified by the coefficient of determination $R^{2}$, where $R^{2} \approx 1$ indicates excellent agreement with the ansatz.}
  \label{fig:typical}
\end{figure}

Fourth, we also studied the typical energy gap $\overline{\ln\delta_\text{e}}$ and the typical bulk-bulk correlations $\overline{\ln C_\text{FM}(r)}$ and $\overline{\ln C_\text{EN}(r)}$ for the QP-Topological Ising transition;
the results are presented in Fig.~\ref{fig:typical}. 
The primary motivation for analyzing these typical quantities is to provide complementary evidence to distinguish the observed QP-Topological Ising transition from the infinite-randomness fixed point (IRFP). 
According to Ref.~\cite{Chepiga2024PRL}, an IRFP is characterized by the typical gap scales as $\overline{\ln \delta_\text{e}} \sim - \sqrt{q}$ and the typical spin-spin correlation decays as a stretched exponential $\overline{\ln C_\text{FM}(r)} \sim - \sqrt{r}$. 
In stark contrast, however, our numerical results exhibit power-law behaviors, scaling as $\overline{\ln \delta_\text{e}} \propto -1.9(1) \ln(q)$, $\overline{\ln C_\text{FM}(r)} \propto - 0.40(1) \ln(r)$, and $\overline{\ln C_\text{EN}(r)} \propto -3.6(1) \ln(r)$. 
As shown in the lower panels of Fig.~\ref{fig:typical}, the numerical data exhibit manifest deviations from the square-root scaling. 
This qualitative discrepancy clearly rules out the IRFP scenario.
Furthermore, we can also exclude the possibility of a clean transition. 
In a conventional clean critical point, the scaling exponents extracted from typical averages should coincide with those from arithmetic averages.
However, our analysis reveals that the logarithmic-scaling coefficients of the typical quantities differ noticeably from the exponents obtained by arithmetic averages, e.g., 2$\Delta_{\sigma}^\text{bulk}$ and 2$\Delta_{\epsilon}^\text{bulk}$. 
This deviation implies that the transition is also not of the clean type. 
Consequently, the combined analysis of the typical gap and correlations confirms that our QP-Topological Ising transition represents a distinct criticality, being neither a clean critical point nor an IRFP.

\begin{figure}[t]
  \centering
  \includegraphics[width=0.5\linewidth]{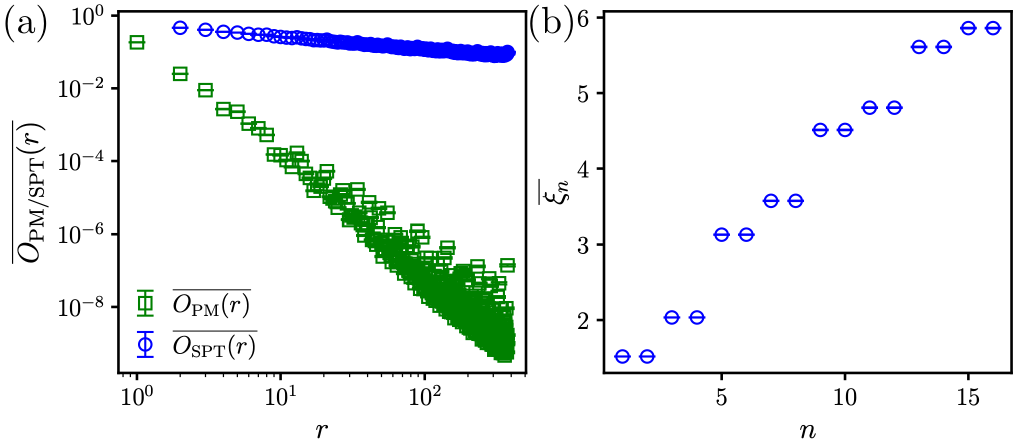}
  \caption{The PM/SPT string order parameter $\overline{O_\text{PM/SPT}(r)}$ and the bulk many-body entanglement spectrum $\overline{\xi_{n}}$ at $\bar{J} = \bar{g} = 1/2, h_{J} = h_{g} = 1$ [the red star in Fig.~1(c) in the main text] in the presence of a small symmetry-preserving perturbation $-h' \sum_{i} \sigma_{i}^{z}$ with $h' = 10^{-3}$. The parameter $q$ is $377$ for (a) and $610$ for (b) under PBC.}
  \label{fig:perturbation}
\end{figure}

Fifth, to examine the robustness of the nontrivial topology of the QP-Topological Ising transition, we add a small symmetry-preserving perturbation term $-h'\sum_{i=1}^{N} \sigma_{i}^{z}$ with $h' = 10^{-3}$ to the QP cluster-Ising model at the critical point $\bar{J} = \bar{g} = 1/2, h_{J} = h_{g} = 1$ [the red star in Fig.~1(c) in the main text]. 
Although the added term can, in principle, alter the critical point, we can still anticipate to observe the topological and critical properties of the transition at the original parameter setting given such a small $h'$.
As shown in Fig.~\ref{fig:perturbation}, the nonlocal string order parameters $\overline{O_\text{PM/SPT}(r)}$ still decay in a power-law way, and $\overline{O_\text{SPT}(r)}$ decays much more slowly than $\overline{O_\text{PM}(r)}$. 
In addition, the bulk many-body entanglement spectrum exhibits a two-fold degeneracy for all levels. 
All these results indicate that the nontrivial topology of the QP-Topological Ising criticality is robust against symmetry-preserving perturbations.

\begin{figure}[t]
  \centering
  \includegraphics[width=\linewidth]{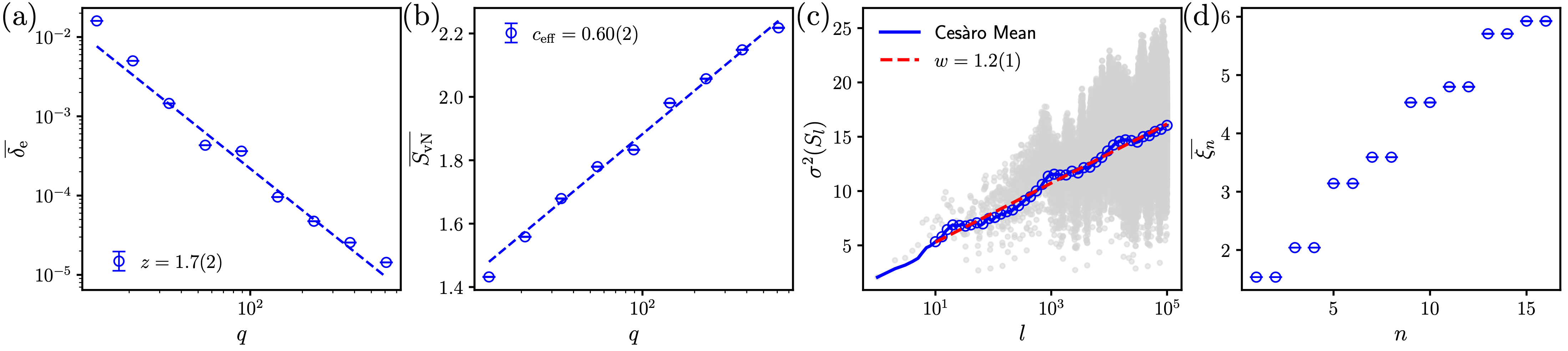}
  \caption{Scaling analysis of the energy gap $\overline{\delta_\text{e}}$, the half-chain entanglement entropy $\overline{S_\text{vN}}$, the variance of wandering $\sigma^{2}(S_{l})$, and the bulk many-body entanglement spectrum $\overline{\xi_{n}}$ at another critical point $\bar{J} = \bar{g} = 1/3, h_{J} = h_{g} = 1$ [the red circle in Fig.~1(c) in the main text] for the QP-Topological Ising transition. The parameter $q$ ranges from $13$ to $610$ for (a) and (b), and $q = 610$ for (d) under PBC. In (c), we have fixed $\phi_{1} = 1/10$ and $\phi_{2} = 1$.}
  \label{fig:down_a}
\end{figure}

\begin{figure}[t]
  \centering
  \includegraphics[width=0.75\linewidth]{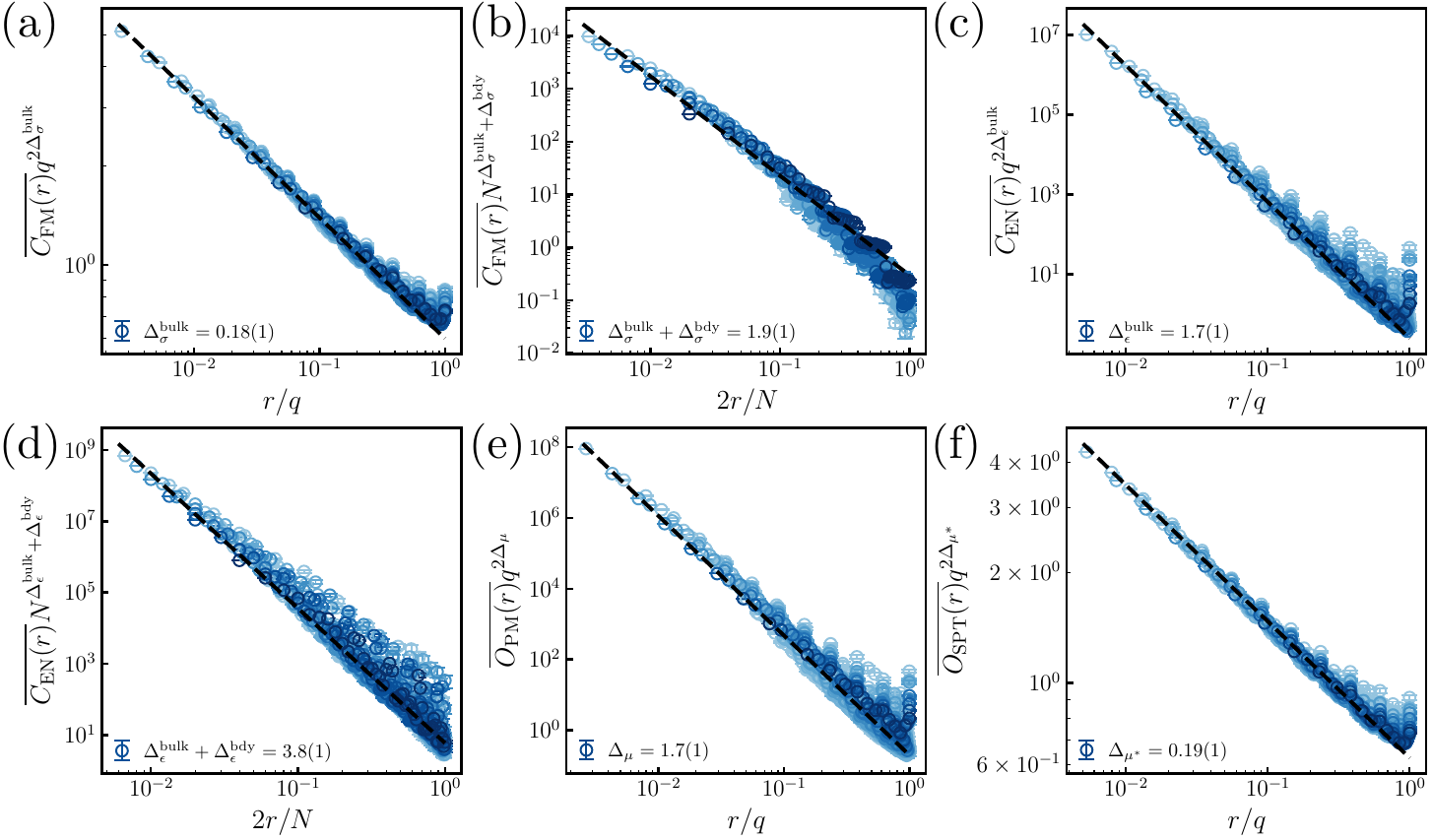}
  \caption{Scaling analysis of various bulk-bulk and bulk-boundary correlations as well as nonlocal string order parameters at another critical point $\bar{J} = \bar{g} = 1/3, h_{J} = h_{g} = 1$ [the red star in Fig.~1(c) in the main text] for the QP-Topological Ising transition. The parameter $q$ ranges from $13$ to $377$ for (a), (c), (e), and (f) under PBC, and $N$ ranges from $100$ to $600$ for (b) and (d) under OBC.}
  \label{fig:down_b}
\end{figure}

Sixth, to verify the universality of our findings along the entire red line in Fig.~1(c) in the main text, we performed the same scaling analysis at another critical point located at $\bar{J} = \bar{g} = 1/3, h_{J} = h_{g} = 1$ [the red circle in Fig.~1(c) in the main text], which also corresponds to the QP-Topological Ising transition. 
As illustrated in Fig.~\ref{fig:down_a} and Fig.~\ref{fig:down_b}, all quantities exhibit critical scaling behaviors. 
The relevant data extracted, including the effective central charge $c_\text{eff}$, the wandering coefficient $w$, and the critical exponents $\Delta_{\sigma/\epsilon}^\text{bulk/bdy}$, are in remarkable agreement with the results obtained at $\bar{J} = \bar{g} = 1/2, h_{J} = h_{g} = 1$ within numerical accuracy. 
This quantitative consistency strongly suggests that the entire red line in the phase diagram belongs to the same universality class, which is characterized by a unique set of critical exponents.

Finally, to support the physical picture that was established at the end of the first section in the End Matter, we further demonstrate that the trivial and topological QP-Ising critical lines exhibit distinct symmetry-enriched realizations of the QP-Ising criticality by considering the following Hamiltonian:
\begin{equation}
  H_\text{tri} = -\sum_{i} J_{i} \sigma_{i}^{x} \sigma_{i+1}^{x} - \sum_{i} f_{i} \sigma_{i}^{z} - \sum_{i} g_{i} \sigma_{i}^{x} \sigma_{i+1}^{z} \sigma_{i+2}^{x} \, ,
  \label{eq:Htri}
\end{equation}
where
\begin{align}
  J_{i} & = \bar{J} + h_{J} \cos\left[ Q(i+1/2)+\phi_{1} \right] \, , \nonumber \\
  g_{i} & = \bar{g} + h_{g} \cos\left[ Qi + \phi_{1} + \phi_{2} \right] \, , \nonumber \\
  f_{i} & = \bar{f} + h_{f} \cos\left[ Qi+\phi_{1}+\phi_{3} \right] \, .
\end{align}
Here, $\sigma_i^{x/z}$ denotes the Pauli matrix at site $i$, $Q$ is the incommensurate modulation wavevector, and $\phi_{1}$, $\phi_{2}$, and $\phi_{3}$ are independent random phases uniformly distributed in $[0, 2\pi)$. This Hamiltonian allows the trivial QP-Ising and topological QP-Ising critical lines to be realized simultaneously within the same phase diagram as shown below. 

\begin{figure}[t]
  \centering
  \includegraphics[width=\linewidth]{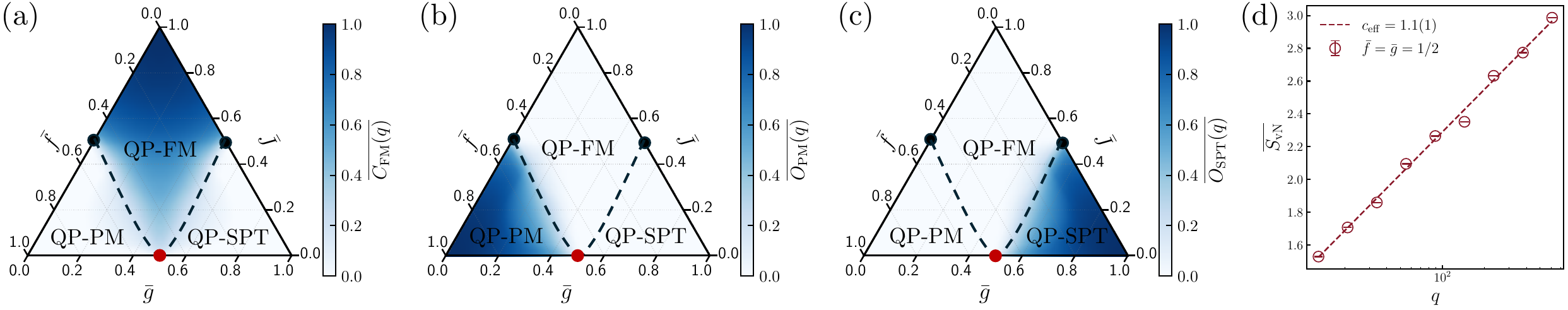}
  \caption{(a) The averaged FM correlation $\overline{C_\text{FM}(q)}$, (b) PM string order parameter $\overline{O_\text{PM}(q)}$, and (c) SPT string order parameter $\overline{O_\text{SPT}(q)}$, shown as a function of $(\bar{J},\bar{f},\bar{g})$ simulated with $(p,q) = (89,55)$ under periodic boundary conditions for the generalized Hamiltonian~\eqref{eq:Htri}. The two dashed lines indicate the trivial and topological QP-Ising critical lines, respectively; the red circle marks the multicritical point that connects them. (d) Logarithmic-law behavior of the half-chain entanglement entropy $\overline{S_{\rm vN}} = \frac{c_\text{eff}}{3} \ln q + {\rm const}$ at the multicritical point $(\bar{J},\bar{f},\bar{g}) = (0,1/2,1/2)$. The effective central charge is estimated $c_\text{eff} = 1.1(1)$\,. The parameter $q$ ranges from $13$ to $610$ under periodic boundary conditions.} 
  \label{fig:universality}
\end{figure}

To systematically map out the global phase diagram, we enforce the normalization condition $\bar{J} + \bar{f} + \bar{g} = 1$, while maintaining the fixed proportion $\bar{J}/h_{J} = \bar{f}/h_{f} = \bar{g}/h_{g} = 1/2$ for the modulation amplitudes.
As shown in Fig.~\ref{fig:universality}(a)-(c), we map out the global phase diagram by calculating the averaged FM correlation, the trivial PM string order parameter, and the SPT string order parameter throughout the entire parameter space. Based on the global behaviors of these observables, we can identify two critical lines: one is a trivial QP-Ising critical line separating the QP-FM and QP-PM phases, while the other is a topological QP-Ising critical line separating the QP-SPT and QP-FM phases. Although we did not locate the precise positions of these critical lines due to the immense computational resources required for performing finite-size scaling analyses, we note that these two phase boundaries are generically expected to be curved rather than strictly straight lines as in the clean limit. This behavior is closely analogous to the disordered case discussed in Ref.~\cite{Duque2021PRB_L}. More importantly, these two QP-Ising critical lines meet at a multicritical point located at $(\bar{J}, \bar{f}, \bar{g}) = (0, 1/2, 1/2)$. At this point, we calculate the entanglement entropy under periodic boundary conditions and find a logarithmic scaling behavior with an effective central charge $c_{\mathrm{eff}}\approx1.1(1)$ [see Fig.~\ref{fig:universality}(d)], which is nearly twice the effective central charge $c_{\mathrm{eff}}\approx0.6$ obtained at the trivial QP-Ising and topological QP-Ising critical points. Therefore, although the two QP-Ising critical lines share the same effective central charge under periodic boundary conditions, within the symmetry-preserving parameter space considered here, the two lines meet at a multicritical point with a different effective central charge. This behavior indicates that the merging of these two QP-Ising critical lines occurs through a multicritical point (i.e., a phase transition between phase transitions), further supporting that they exhibit distinct symmetry-enriched realizations of the same QP-Ising criticality.

\section{Numerical results for Silver ratio}
\label{S3}

\begin{figure}[t]
  \centering
  \includegraphics[width=\linewidth]{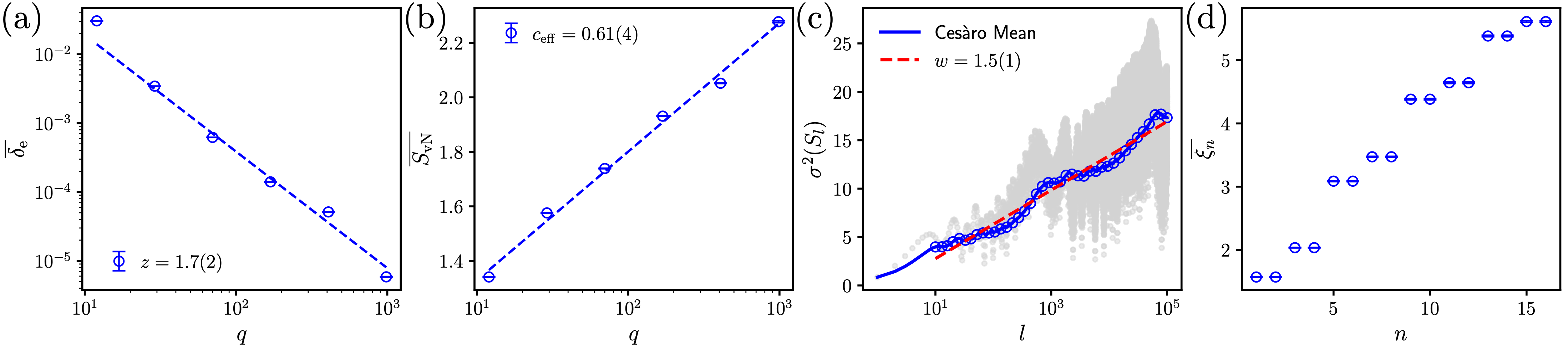}
  \caption{Scaling analysis of the averaged energy gap $\overline{\delta_\text{e}}$, the half-chain entanglement entropy $\overline{S_\text{vN}}$, the variance of wandering $\sigma^{2}(S_{l})$, and the bulk entanglement spectrum $\overline{\xi_{n}}$ at $\bar{J} = \bar{g} = 1/2, h_{J} = h_{g} = 1$ [the red star in Fig.~1(c) in the main text]) for the Silver ratio $Q/(2\pi) = \tau_\text{S}$. The parameter $q$ ranges from $12$ to $985$ for (a) and (b), and $q = 985$ for (d) under PBC. In (c), we have fixed $\phi_{1} = 1/10$ and $\phi_{2} = 1$.}
  \label{fig:silver_ratio_a}
\end{figure}

\begin{figure}[t]
  \centering
  \includegraphics[width=0.75\linewidth]{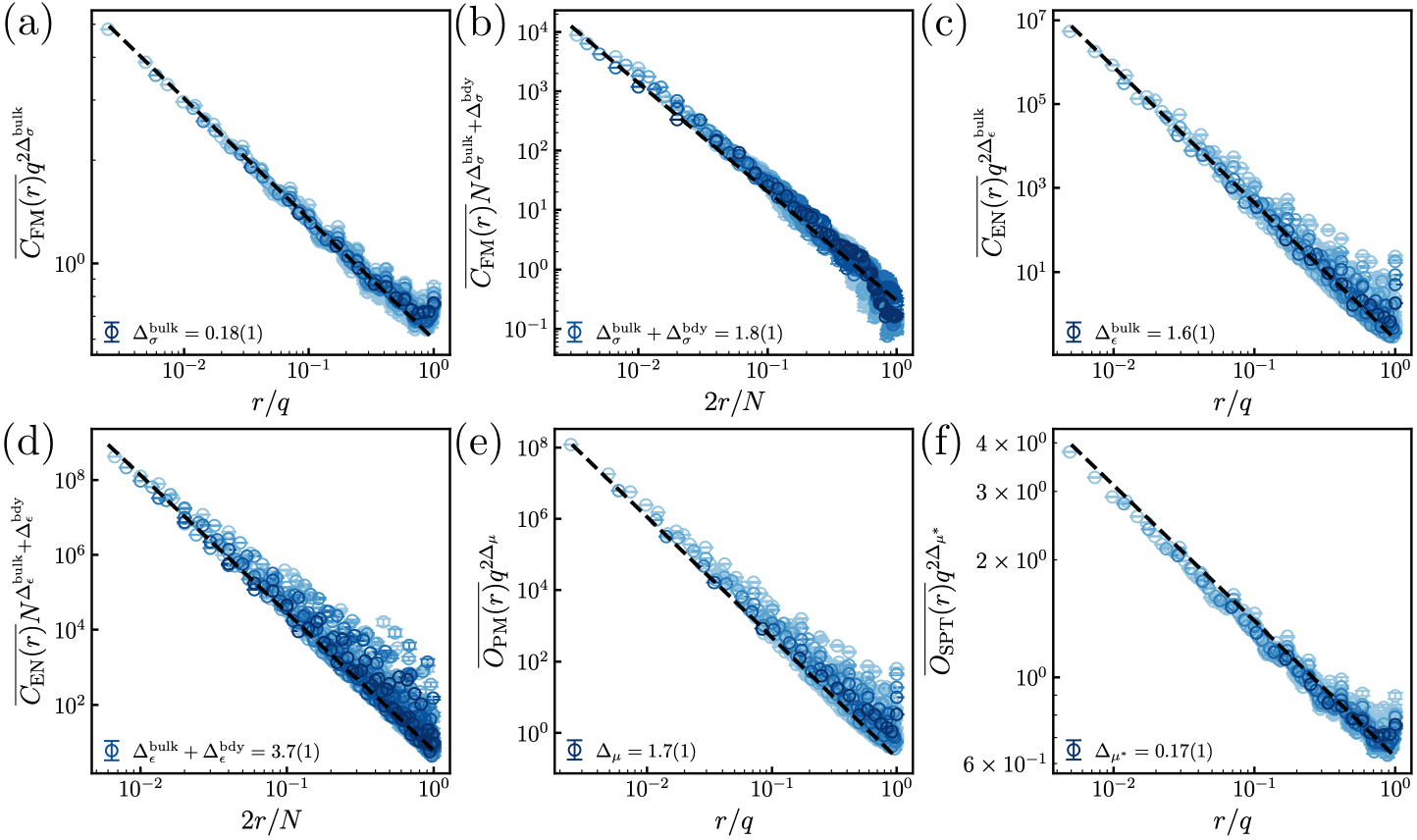}
  \caption{Scaling analysis of various bulk-bulk and bulk-boundary correlations as well as nonlocal string order parameters at $\bar{J} = \bar{g} = 1/2, h_{J} = h_{g} = 1$ [the red star in Fig.~1(c) in the main text] for the Silver ratio $Q/(2\pi) = \tau_\text{S}$. The parameter $q$ ranges from $12$ to $408$ for (a), (c), (e), and (f) under PBC, and $N$ ranges from $100$ to $600$ for (b) and (d) under OBC.}
  \label{fig:silver_ratio_b}
\end{figure}

To show that our results are robust and not restricted to the specific choice of the Golden ratio $\tau_\text{G}$, we extend our analysis in this section to a broad class of irrational numbers by considering the Silver ratio, $\tau_\text{S} = 1 + \sqrt{2}$.
Similarly, $\tau_\text{S}$ can be rationally approximated by the ratio of consecutive Pell numbers, i.e., $\tau_\text{S} \approx P_{n+1}/P_{n}$. 
The Pell numbers are defined recursively by $P_{n} = 2P_{n-1} + P_{n-2}$ with initial conditions $P_{0} = 0$ and $P_{1} = 1$, generating the series $0, 1, 2, 5, 12, 29, 70, 169, 408, 985, \dots$

We focus on the same critical point $\bar{J} = \bar{g} = 1/2, h_{J} = h_{g} = 1$ [the red star in Fig.~1(c) in the main text] for the QP-Topological Ising transition, but now with the modulation wave vector set to $Q/(2\pi) = \tau_\text{S}$.
The numerical results are presented in Fig.~\ref{fig:silver_ratio_a} and Fig.~\ref{fig:silver_ratio_b}. 
We observe that all physical behaviors are qualitatively similar to those obtained with $\tau_\text{G}$. 
Remarkably, the data extracted for the criticality, such as the effective central charge and critical exponents, are found to be close to the case with $\tau_\text{G}$, providing strong evidence for the universality of the QP-Topological Ising criticality across QP-modulations with different irrational numbers.

\section{Numerical results for the remaining two phase boundaries}
\label{S4}

\begin{figure}[t]
  \centering
  \includegraphics[width=\linewidth]{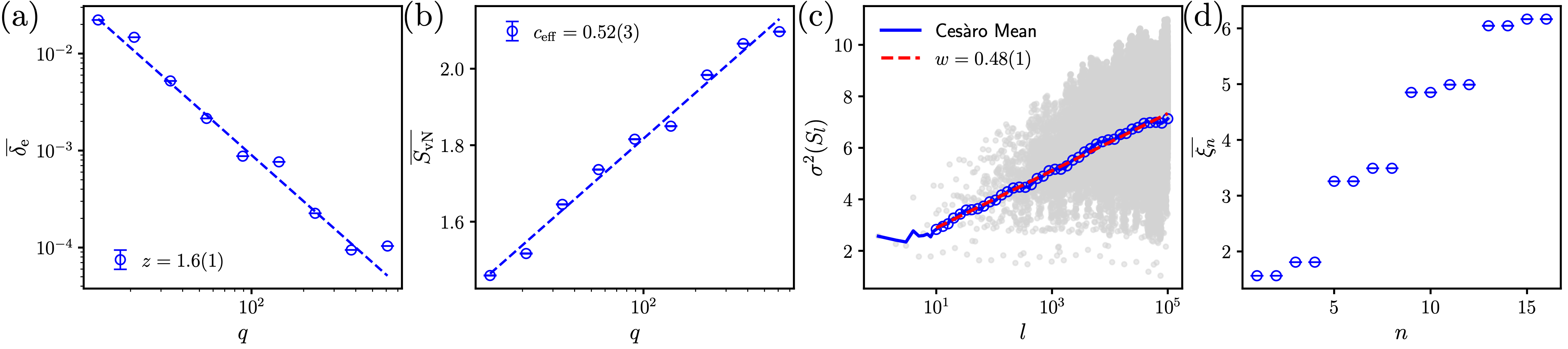}
  \caption{Scaling analysis of the energy gap $\overline{\delta_\text{e}}$, the half-chain entanglement entropy $\overline{S_\text{vN}}$, the variance of wandering $\sigma^{2}(S_{l})$, and the bulk many-body entanglement spectrum $\overline{\xi_{n}}$ at a representative point $\bar{J} = \bar{g} = 5/8, h_{J} = 1/2, h_{g} = 1$ [the up-pointing triangular in Fig.~1(c) in the main text] on the critical line separating the FM and QP-SPT phases. The parameter $q$ ranges from $13$ to $610$ for (a) and (b), and $q = 610$ for (d) under PBC. In (c), we have fixed $\phi_{1} = 1/10$ and $\phi_{2} = 1$.}
  \label{fig:left_a}
\end{figure}

\begin{figure}[t]
  \centering
  \includegraphics[width=0.75\linewidth]{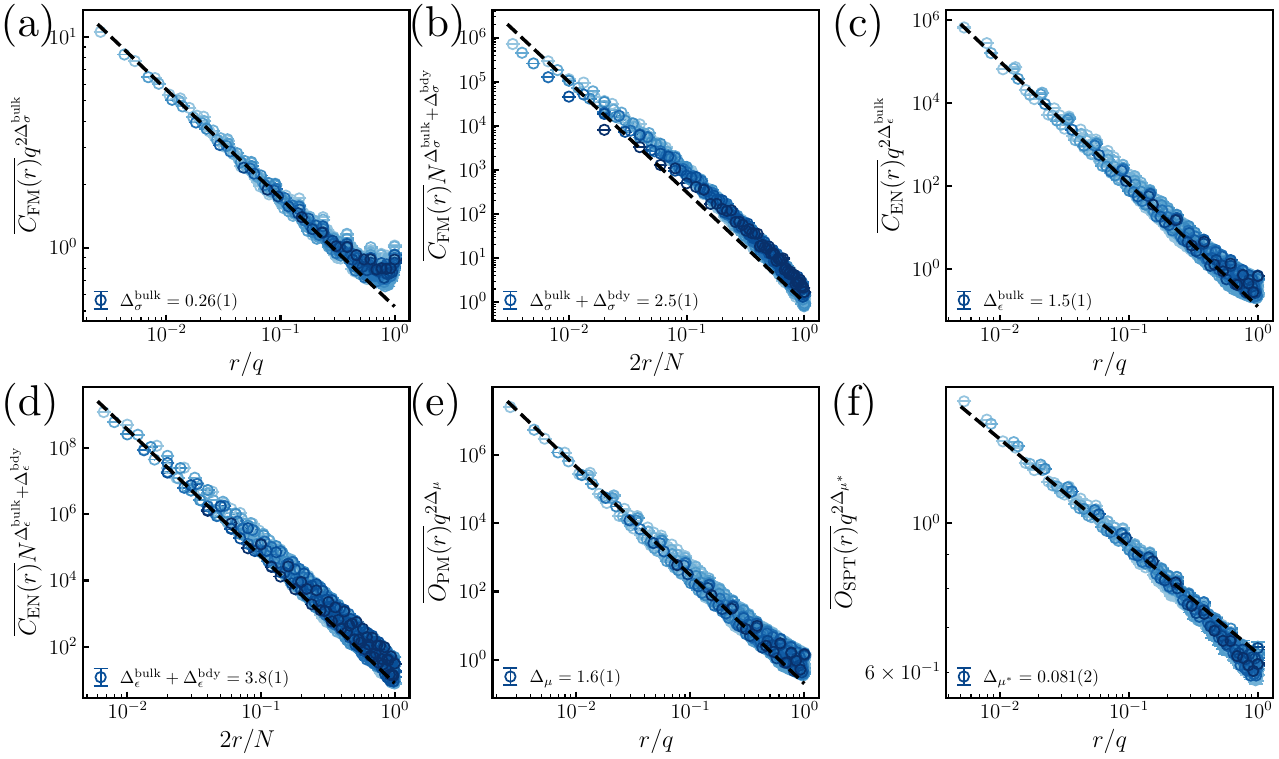}
  \caption{Scaling analysis of various bulk-bulk and bulk-boundary correlations at a representative point $\bar{J} = \bar{g} = 5/8, h_{J} = 1/2, h_{g} = 1$ [the up-pointing triangular in Fig.~1(c) in the main text] on the critical line separating the FM and QP-SPT phases. The parameter $q$ ranges from $13$ to $377$ in (a), (c), (e), and (f) under PBC, and $N$ ranges from $100$ to $600$ in (b) and (d) under OBC.}
  \label{fig:left_b}
\end{figure}
 
\begin{figure}[t]
  \centering
  \includegraphics[width=\linewidth]{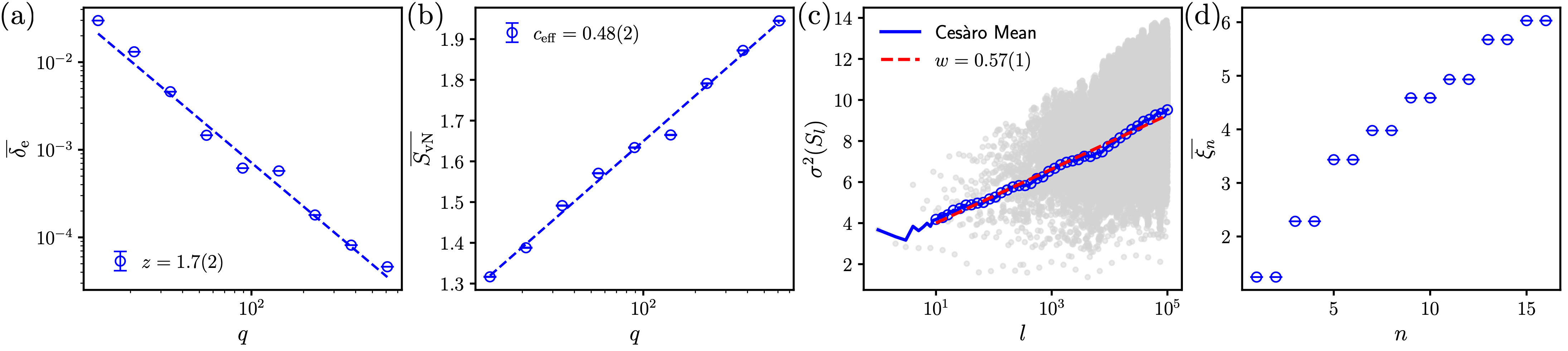}
  \caption{Scaling analysis of the energy gap $\overline{\delta_\text{e}}$, the half-chain entanglement entropy $\overline{S_\text{vN}}$, the variance of wandering $\sigma^{2}(S_{l})$, and the bulk many-body entanglement spectrum $\overline{\xi_{n}}$ at a representative point $\bar{J} = \bar{g} = 13/12, h_{J} = 3/2, h_{g} = 1$ [the down-pointing triangular in Fig.~1(c) in the main text] on the critical line separating the SPT and QP-FM phases. The parameter $q$ ranges from $13$ to $610$ for (a) and (b), and $q = 610$ for (d) under PBC. In (c), we have fixed $\phi_{1} = 1/10$ and $\phi_{2} = 1$.}
  \label{fig:right_a}
\end{figure}

\begin{figure}[t]
  \centering
  \includegraphics[width=0.75\linewidth]{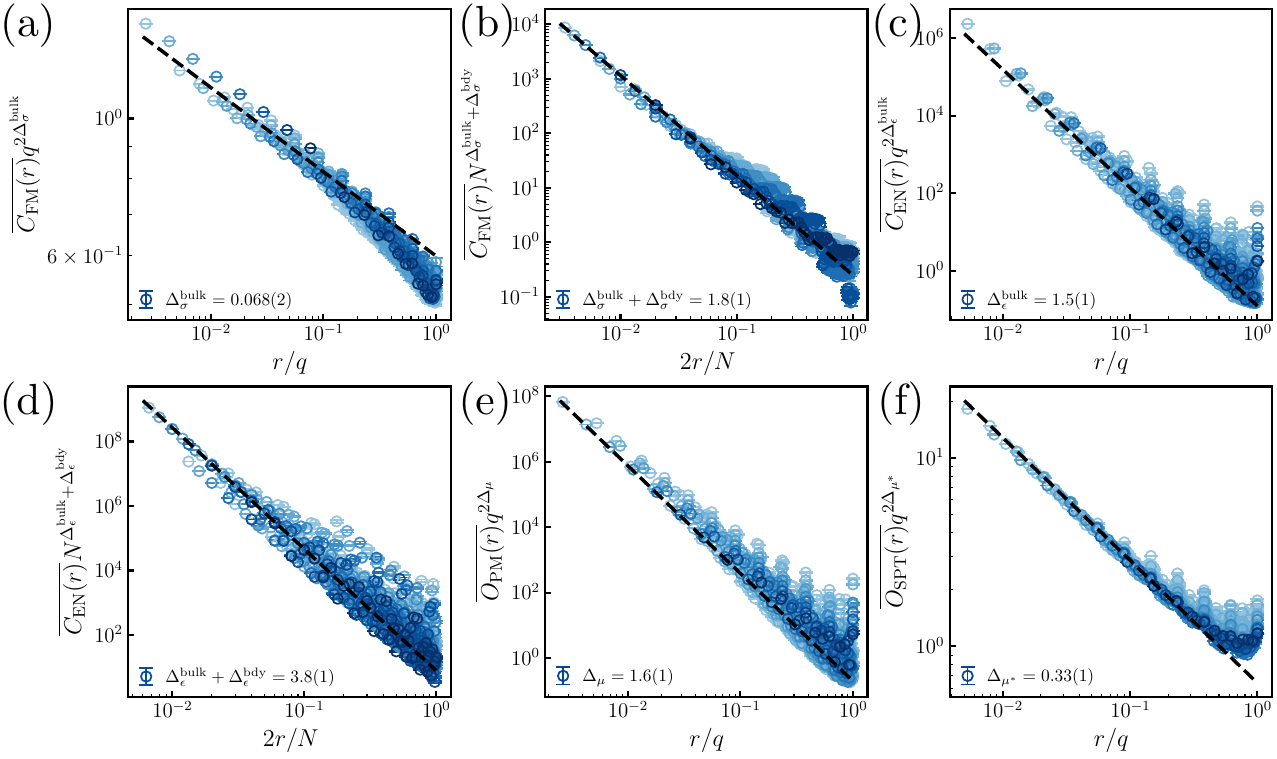}
  \caption{Scaling analysis of various bulk-bulk and bulk-boundary correlations at a representative point $\bar{J} = \bar{g} = 13/12, h_{J} = 3/2, h_{g} = 1$ [the down-pointing triangular in Fig.~1(c) in the main tex] on the critical line separating the SPT and QP-FM phases. The parameter $q$ ranges from $13$ to $377$ in (a), (c), (e), and (f) under PBC, and $N$ ranges from $100$ to $600$ in (b) and (d) under OBC.}
  \label{fig:right_b}
\end{figure}

In this section, we extend our analysis to the remaining two phase boundaries depicted in Fig.~1(c) of the main text. 
We perform finite-size scaling analyses at two representative points: (i) $\bar{J} = \bar{g} = 5/8, h_{J} = 1/2, h_{g} = 1$ for the FM to QP-SPT transition (results are shown in Fig.~\ref{fig:left_a} and Fig.~\ref{fig:left_b}); and (ii) $\bar{J} = \bar{g} = 13/12, h_{J} = 3/2, h_{g} = 1$ for the SPT to QP-FM transition (results are shown in Fig.~\ref{fig:right_a} and Fig.~\ref{fig:right_b}).
Our results demonstrate that both phase transitions exhibit critical exponents that are distinct from those of the clean Ising and IRFP.
Furthermore, at the critical point of the QP cluster-Ising chain, the nonlocal string operator $\overline{O_{\mathrm{SPT}}(r)}$ decays more slowly than $\overline{O_{\mathrm{PM}}(r)}$. 
By contrast, at the corresponding critical point of the QP transverse-field Ising chain, $\overline{O_{\mathrm{PM}}(r)}$ must exhibits a slower decay behavior than $\overline{O_{\mathrm{SPT}}(r)}$, since the two models are related by an SPT entangler~\cite{Verresen2021PRX} under PBC, which effectively interchanges the PM and SPT string order parameters. 
These behaviors imply the topological distinction between the critical points of the QP cluster-Ising and QP transverse-field Ising chains, where the nontrivial topology also directly manifests as a twofold degeneracy in the many-body entanglement spectrum.

\bibliography{main.bib}